# Long-range nontopological edge currents in charge-neutral graphene

A. Aharon-Steinberg[1,†], A. Marguerite[1,†], D. J. Perello[2], K. Bagani[1], T. Holder[1], Y. Myasoedov[1], L. S. Levitov[3], A. K. Geim[2], and E. Zeldov[1*]

Van der Waals heterostructures display a rich variety of unique electronic properties. To identify novel transport mechanisms, nonlocal measurements have been widely used, wherein a voltage is measured at contacts placed far away from the expected classical flow of charge carriers. This approach was employed in search of dissipationless spin and valley transport [1–13], topological charge-neutral currents [14–18], hydrodynamic flows [19] and helical edge modes [20–23]. Monolayer [1–8,10,14,16,21–26], bilayer [12,15,17,20,27], and few-layer [28] graphene, transition-metal dichalcogenides [9,11], and moiré superlattices [13,14,16–18] were found to display pronounced nonlocal effects. However, the origin of these effects is hotly debated [5,15,25,29–31]. Graphene, in particular, exhibits giant nonlocality at charge neutrality [1,7,21–26], a prominent behavior that attracted competing explanations. Utilizing superconducting quantum interference device on a tip (SQUID-on-tip) for nanoscale thermal and scanning gate imaging [32], we demonstrate that the commonly-occurring charge accumulation at graphene edges [30,33–38] leads to giant nonlocality, producing narrow conductive channels that support long-range currents. Unexpectedly, while the edge conductance has little impact on the current flow in zero magnetic field, it leads to field-induced decoupling between edge and bulk transport at moderate fields. The resulting giant nonlocality both at charge neutrality and away from it produces exotic flow patterns in which charges can flow against the global electric field. We have visualized surprisingly intricate patterns of nonlocal currents, which are sensitive to edge disorder. The observed one-dimensional edge transport, being generic and nontopological, is expected to support nonlocal transport in many electronic systems, offering insight into numerous controversies in the literature and linking them to long-range guided electronic states at system edges.

________________________________

[1]Department of Condensed Matter Physics, Weizmann Institute of Science, Rehovot 7610001, Israel

[2]National Graphene Institute and School of Physics and Astronomy, The University of Manchester, Manchester M13 9PL, UK

[3]Department of Physics, Massachusetts Institute of Technology, Cambridge, Massachusetts 02139, USA

[†]These authors contributed equally to this work

[*]eli.zeldov@weizmann.ac.il

The guiding principle of nonlocal measurements is based on the property of systems with homogeneous local ohmic resistivity to produce a response at remote contacts that rapidly decays with distance. In a long bar (Fig. 1a), the nonlocal resistance decays exponentially with the distance $x$ from the current-injecting contacts, $R_{NL} = V_{NL}/I_0 \cong \frac{4}{\pi}\rho_{xx}e^{-|x|/\lambda}$, where $\lambda = W/\pi$, $W$ is the bar width, $\rho_{xx}$ is resistivity, $V_{NL}$ is the nonlocal voltage, and $I_0$ is the applied current [1]. This relation is remarkably robust and independent of the carrier type, scattering mechanisms, temperature, or magnetic field. Any deviation from it thus provides a very sensitive means for detecting unconventional transport. In contrast to local measurements in which small deviations from conventional ohmic behavior will cause respectively small deviations in the measured signal, in nonlocal configuration the ohmic contribution vanishes exponentially thus enhancing the relative sensitivity to unconventional mechanisms by orders of magnitude. Hence the nonlocal transport measurements have turned into a method of choice for charting novel transport phenomena in 2D materials [1–29,39–42].

However, linking the observed nonlocal response to underlying physics has proven to be a challenge. This is exemplified most clearly by the giant nonlocality observed in graphene near the charge neutrality point (CNP) [1,7,10,21–26]. At zero magnetic field the measured $R_{NL}$ is found to somewhat exceed the expected, albeit exponentially suppressed, ohmic response. However, $R_{NL}$ sharply increases upon applying a perpendicular magnetic field, exceeding the expected values by orders of magnitude and persisting up to room temperature. Despite the numerous studies and a wide palette of considered mechanisms, the underlying origin of the nonlocality remains highly controversial [1,5,15,25,29–31].

Here we resolve this conundrum utilizing SQUID-on-tip (SOT) thermal imaging and scanning gate microscopy [32,38,43] combined with theoretical analysis. Namely, we demonstrate that at elevated magnetic fields the long-range edge currents arise naturally from the presence of charge accumulation along graphene edges, an effect inferred in previous studies [30,33–38]. Our measurements reveal a highly intricate and counterintuitive transport behavior due to these currents, supporting the picture of electronic one-dimensional states of nontopological origin guided along edges. In magnetic field, these edge states are found to induce an "upstream" flow in which a substantial component of the bulk current flows opposite to the electric field direction. The ubiquitous nature of the nonlocal response as well as its giant enhancement in moderate magnetic fields follow naturally within this framework, shedding light on, and offering an alternative explanation for, a wide range of disputed observations.

**Nonlocal transport measurements**

Graphene encapsulated in hBN was patterned into a Hall bar structure (Fig. 1a and Methods) and studied at 4.2 K in the presence of moderate out-of-plane magnetic fields $B \leq 5$ T employing three types of measurements: transport, scanning gate microscopy, and nanoscale thermal imaging (Methods). Similarly to the previous reports [1,7,21–26], in the vicinity of CNP (back gate voltage $V_{bg} = 0$ V) our graphene device shows $R_{NL}$ which grows rapidly with $B$ (Fig. 1c). In the ohmic response regime, the prefactor of $R_{NL} \cong \frac{4}{\pi}\rho_{xx}e^{-\pi|x|/W}$ is proportional to $\rho_{xx}$ that decreases by about two orders of magnitude upon increasing the carrier concentration with $|V_{bg}|$. Since here we are interested in understanding the mechanism governing the nonlocal decay length $\lambda$, it is more informative to inspect the normalized $\mathcal{R}_{NL} = R_{NL}/R_{2p} = V_{NL}/V_0 = \mathcal{V}_{NL} \cong e^{-|x|/\lambda}$, where $R_{2p} = V_0/I_0$ is the two-probe resistance and $V_0$ is the bias voltage (corrected for small contact resistance, see Methods). The $\mathcal{R}_{NL}$ in Fig. 1d reveals that upon normalizing out the $V_{bg}$ dependence of $\rho_{xx}$, the nonlocality is present over a wide range of $V_{bg}$. Moreover, the logarithmic plot in Fig. 1e shows that the nonlocality at elevated fields is giant, exceeding the ohmic $\mathcal{R}_{NL} \cong \frac{4}{\pi}\frac{\rho_{xx}}{R_{2p}}e^{-\pi x/W} \cong 4.6\times10^{-5}$ by orders of magnitude (dashed horizontal line for our geometry with $x/W \cong 2.85$, see Supplementary Information).

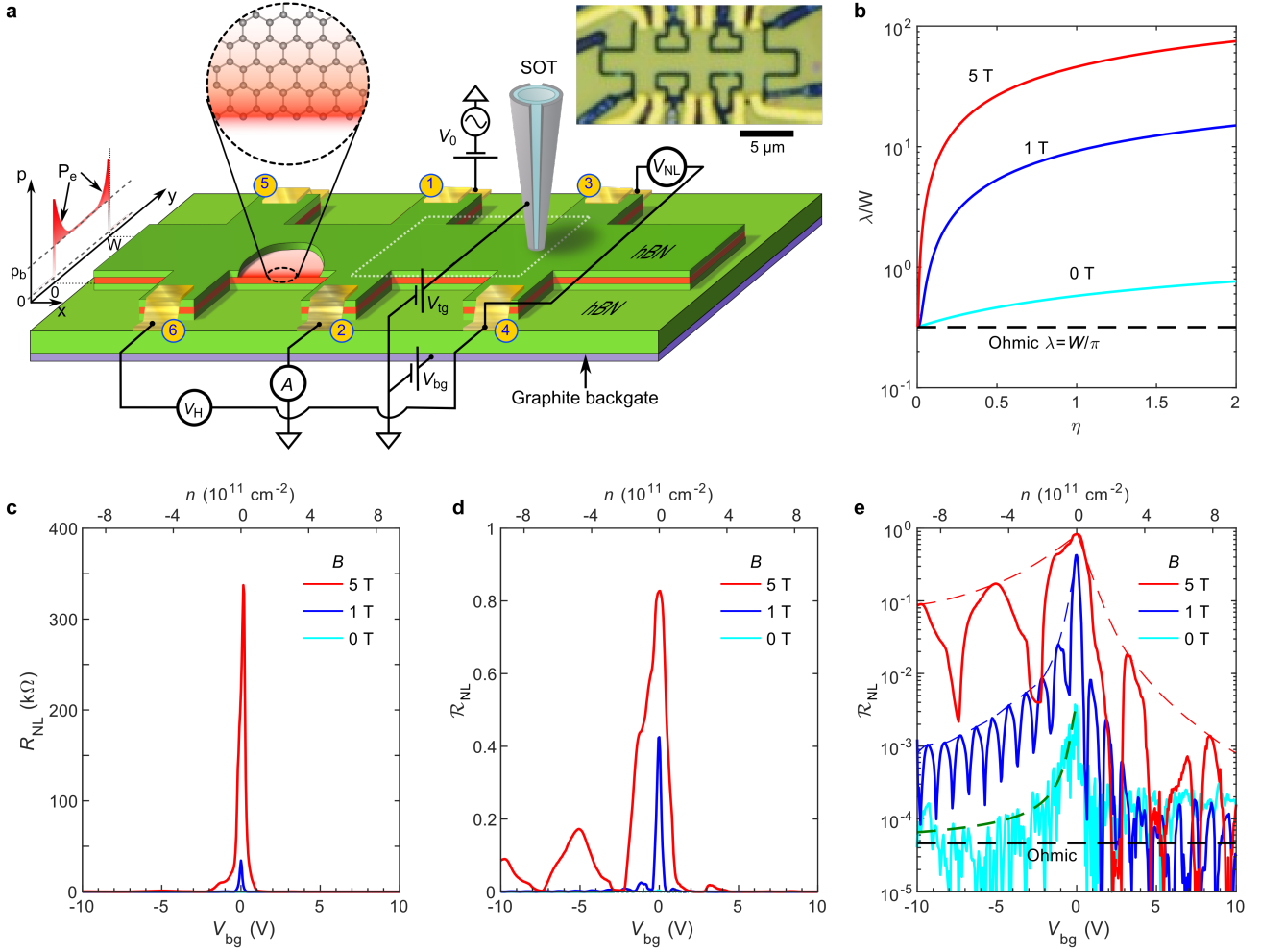

**Fig. 1**. **Nonlocal transport characteristics. a**, Schematic layout of the graphene sample with contact configuration and scanning SOT. The plot on the left shows schematically the hole carrier density distribution, $p$, across the sample width, $W$, with bulk density, $p_b$, and accumulation along the two edges with an excess hole line density, $P_e$ (red). The white dotted contour marks schematically the area imaged in Fig. 3. Inset: Optical image of the hBN encapsulated graphene sample patterned into a Hall bar shape (dark yellow) with gold contacts (bright yellow). **b**, The calculated exponential decay length $\lambda$ of the nonlocal resistance vs. the edge accumulation ratio, $\eta = 2P_e/Wp_b$. In high mobility samples (here $\mu = 2.6\times10^5$ cm$^2$V$^{-1}$s$^{-1}$) the nonlocality increases drastically with $B$ with $\lambda \cong \frac{\eta \mu B W}{2\sqrt{\eta+1}}$. **c**, The nonlocal resistance $R_{NL}$ vs. back gate voltage $V_{bg}$ at $B=0$, 1, and 5 T. **d**, The same $R_{NL}$ normalized by two-probe resistance, $\mathcal{R}_{NL} = R_{NL}/R_{2p}$. **e**, $\mathcal{R}_{NL}$ on a logarithmic scale revealing the giant nonlocality relative to the ohmic case (black dashed line). The nonlocality is maximal at CNP but remains very large even at high carrier densities in particular for $p$ doping ($V_{bg} < 0$). The green dashed line shows the analytic solution of $\mathcal{R}_{NL}(V_{bg})$ with $P_e = 1.8\times10^8$ m$^{-1}$ at $B=0$ T for an infinite sample. The red and blue dashed lines are guide-to-the-eye envelopes of $\mathcal{R}_{NL}$ at $B=1$ and 5 T. See Supplementary Information for parameters.

Figure 1e demonstrates a number of important features. *i*) A finite nonlocality above the ohmic value is present even at zero field near CNP. *ii*) Near CNP $\mathcal{R}_{NL}$ grows rapidly with field reaching a four orders of magnitude enhancement over the ohmic value. *iii*) At high fields near CNP, $\mathcal{R}_{NL}$ approaches a value close to 1, namely the nonlocal voltage $V_{NL}$ at the remote contacts becomes comparable to the applied voltage $V_0$, rather than being exponentially small. *iv*) The rate at which $\mathcal{R}_{NL}$ decreases with $|V_{bg}|$ (dashed guide-to-the-eye envelope curves) slows down significantly with increasing field. At $B = 5$ T and $V_{bg} = -10$ V, $\mathcal{R}_{NL}$ remains giant exceeding the ohmic value by more than three orders of magnitude. *v*) The nonlocality is asymmetric with respect to hole and electron doping, displaying a faster decay with positive $V_{bg}$. *vi*) By comparing $\mathcal{R}_{NL}(V_{bg})$ at various fields (Extended Data Figs. 1c,d) it becomes apparent that the nonlocal mechanism determines the envelopes of $\mathcal{R}_{NL}(V_{bg})$ (dashed curves) which decrease with $|V_{bg}|$ and increase with $B$, but are continuous and smooth as a function of both. *vii*) The periodic dips in $\mathcal{R}_{NL}$ occur at $V_{bg}$ values corresponding to the integer quantum Hall (QH) effect plateaus (Extended Data Figs. 1a,b). In the QH plateaus any four-probe measurement of $R_{xx}$ in either local or nonlocal configuration should vanish due to vanishing $\rho_{xx}$, while $R_{2p}$ remains finite, leading to the observed vanishing of $\mathcal{R}_{NL}$. We emphasize, however, that in the QH plateau transition regions, where $\rho_{xx}$ is finite, $\mathcal{R}_{NL}(V_{bg}, B)$ should not exceed the ohmic value (black dashed curve) within the ohmic description. To resolve the mechanism of the nonlocality, we therefore focus hereafter on the regions near the CNP as well as near the local maxima of $\mathcal{R}_{NL}(V_{bg})$. The corresponding behavior in the quantized QH state at the minima of $\mathcal{R}_{NL}(V_{bg})$ is expected to be more involved and is outside the scope of this paper.

**Thermal imaging of nonlocal dissipation**

Understanding dissipation in the nonlocal regime is of keen interest because it will help to delineate between the nonlocality arising due to dissipationless currents and other mechanisms. To characterize the nonlocal aspects of dissipation we employ the SOT as a nanothermometer with µK sensitivity scanning at a height of about 100 nm above the sample surface (Methods) [32]. In the ohmic regime the currents that flow in the remote regions are exponentially small which should lead to negligible dissipation and heating there. The scanning SOT thermal imaging in Fig. 2c indeed shows that at $B = 0$ T and $V_{bg} = 1$ V, at which no nonlocal contribution to $\mathcal{R}_{NL}$ is observed in Fig. 1e, the heat dissipation occurs predominantly in the central part of the sample as expected in the case of local ohmic transport. At charge neutrality, however, where an enhanced $\mathcal{R}_{NL}$ is observed in Fig. 1e, the temperature profile is distinctively more extended along the $x$ direction with heat dissipation protruding into the right and left "nonlocal" arms marked by *R* and *L* respectively in Fig. 2b. A similar, but weaker spread of the thermal signal is visible in Fig. 2a at $V_{bg} = -1$ V where a small but finite nonlocal contribution to $\mathcal{R}_{NL}$ is present in Fig. 1e (see Supplementary Video 1 for range of $V_{bg}$). Note that in the ohmic regime at $B = 0$ T, the current and electric field distribution functions in the sample should be independent of $\rho_{xx}$ and hence varying $V_{bg}$ may only change the overall intensity of the temperature profile, but not its spatial dependence (neglecting dissipation at contacts). The thermal images therefore indicate that the observed enhancement in $\mathcal{R}_{NL}$ is associated with a dissipative process.

When $B$ is increased to 1 T the nonlocal temperature distribution at CNP becomes more extended (Fig. 2d) and by 5 T the thermal signal expands throughout the entire sample (Fig. 2f). In the ohmic regime in addition to being independent of $V_{bg}$, the current distribution within the sample should also be field independent [1]. The striking spread of the thermal signal across the sample at elevated $B$ (see Supplementary Video 2) thus reveals an unconventional dissipation mechanism. Moreover, upon applying a positive potential $V_{tg}$ to the SOT (Fig. 1a) the thermal signal is observed to be enhanced further (Figs. 2e,g and Supplementary Video 3), but predominantly along the edges of the sample, forming a highly irregular profile. This finding points towards the dominant role played by the edges in the dissipation process.

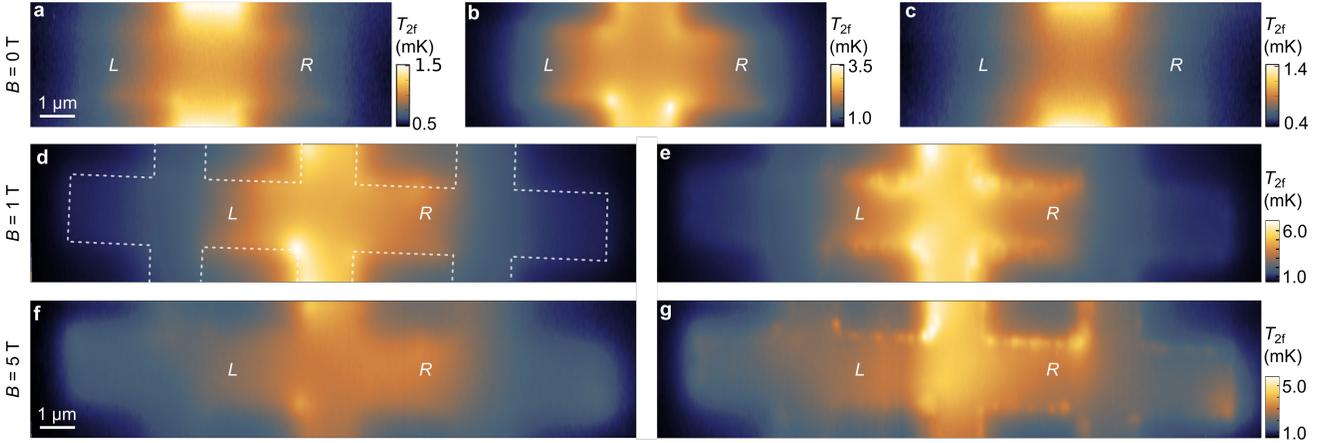

**Fig. 2. Scanning SOT thermal imaging of the graphene sample at 4.2 K**. An *ac* voltage $V_0$ at frequency $f = 66.66$ Hz is applied to the central top contact, 1, while the bottom contact, 2, is grounded (see Fig. 1a). The resulting thermal image, $T_{2f}$, is acquired by the scanning SOT at frequency $2f$. The power applied to the sample, $V_0 I_0 = 15$ nW, is kept constant upon varying $V_{bg}$ and the magnetic field $B$. **a-c**, Temperature maps $T_{2f}$ of the central part of the sample at $B = 0$ T and $V_{tg} = 0$ V at $V_{bg} = -1$ V (**a**), 0 V (**b**), and 1 V (**c**). At CNP (**b**), a moderate spread of the thermal signal into the right (*R*) and left (*L*) arms is visible relative to **c**. See Supplementary Video 1 for range of $V_{bg}$. **d-e**, $T_{2f}$ images of the full sample near CNP ($V_{bg} = -0.05$ V) at $B = 1$ T acquired with $V_{tg} = 0$ V (**d**) and $V_{tg} = 8$ V applied to the SOT (**e**). The spread of the dissipation into the *R* and *L* arms (**d**) is enhanced relative to **b** and the dissipation along the edges becomes prominent in the presence of $V_{tg}$ (**e**). The white dotted contour in **d** outlines the sample edges. **f-g**, $T_{2f}$ images at $V_{bg} = 0$ V at $B = 5$ T with $V_{tg} = 0$ V (**f**) and $V_{tg} = 8$ V (**g**), revealing the expansion of the dissipation all the way to the far ends of the *R* and *L* arms (**f,g**). The effects of edge disorder are clearly visible in **g**. See Supplementary Videos 2 and 3 for range of $V_{bg}$ and Supplementary Information for parameters.

**Scanning gate microscopy of nonlocal transport**

The indirect signatures of edge currents, obtained from measuring dissipation, can be transformed into a direct and unambiguous evidence by using the SOT probe in the scanning gate mode. To that end, we apply a potential $V_{tg}$ to the SOT and measure $\mathcal{V}_{NL} = V_{NL}/V_0$ vs. tip position at a fixed sample bias voltage $V_0$. At negative $V_{bg}$, application of a positive $V_{tg}$ causes depletion of the hole carrier density, $p$, under the tip, thus reducing the local conductivity and perturbing the local current flow. Figure 3a shows that this has a profound effect on $\mathcal{V}_{NL}$. When the tip is outside the sample the measured nonlocality at $B = 5$ T and $V_{bg} = -0.66$ V is giant with $\mathcal{V}_{NL} = 0.26$, that is over three orders of magnitude above the ohmic value (see Supplementary Video 4 for other $V_{bg}$ values). Scanning the tip over the interior of the sample and depleting the local $p$ in the bulk has essentially no effect on $\mathcal{V}_{NL}$. However, when the tip is located above the edges of the sample, $\mathcal{V}_{NL}$ drops sharply down to values close to zero, quenching the giant nonlocality. The suppression of $\mathcal{V}_{NL}$ along the edges is highly nonuniform with some locations displaying an almost full suppression while others show almost no effect. Moreover, at more negative $V_{bg}$ values, an asymmetry between the edges develops with strongly diminished scanning gate signal along the bottom-right edge as shown in Supplementary Video 4 and Extended Data Fig. 4a for $V_{bg} = -1.6$ V.

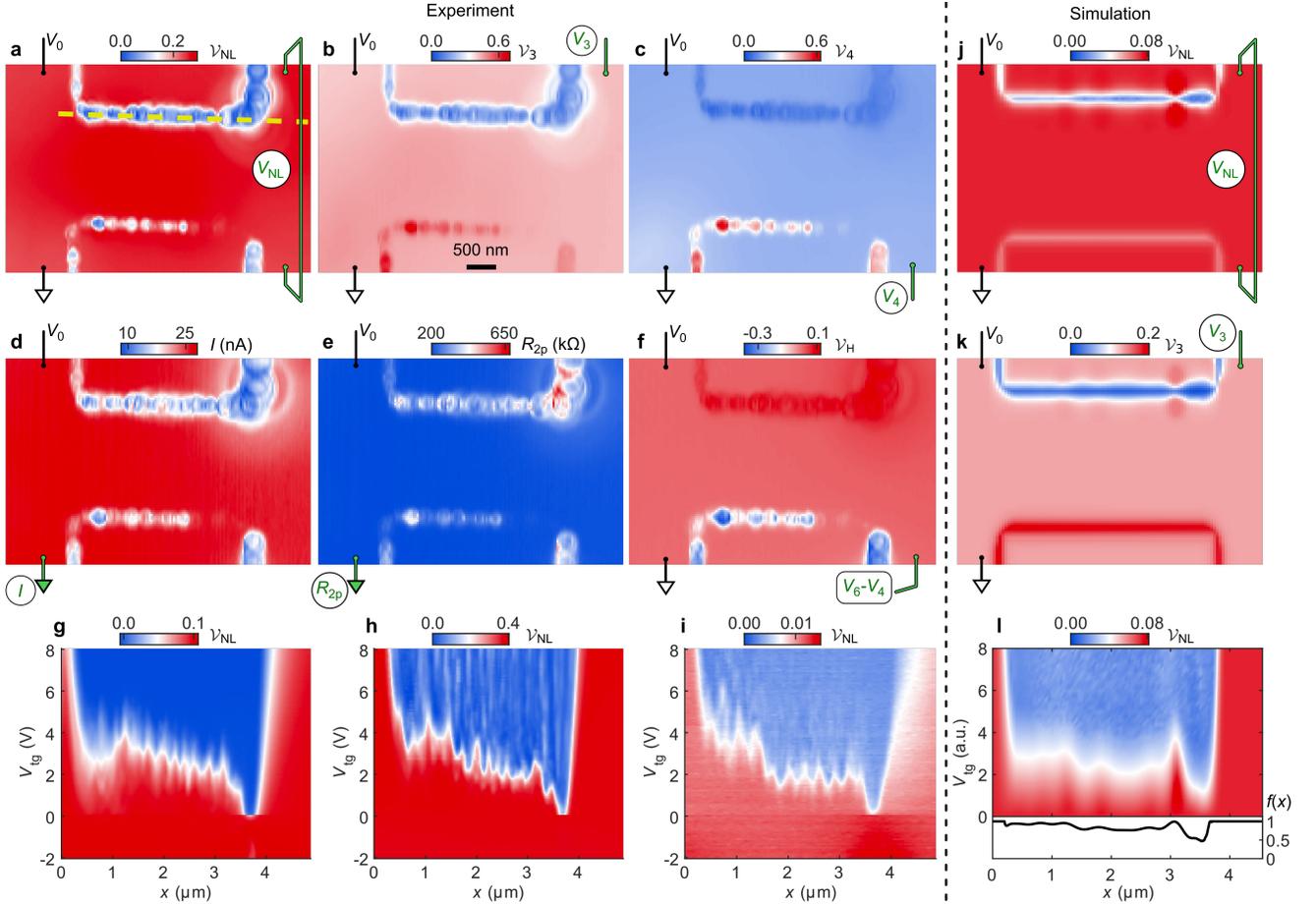

**Fig. 3. Scanning gate microscopy at $B = 5$ T**. The graphene device is biased with a voltage $V_0 = 5.5$ mV and the SOT is scanned over the right-hand-side of the sample (white dotted area in Fig. 1a) while the current and the potentials at various contacts are measured. **a-c**, Images of $\mathcal{V}_{NL} = (V_3 - V_4)/V_0$ (**a**), $\mathcal{V}_3 = V_3/V_0$ (**b**), and $\mathcal{V}_4$ (**c**) at $V_{bg} = -0.66$ V and $V_{tg} = 8$ V (see Supplementary Video 4 for range of $V_{bg}$ values). **d-f**, Images of $I$ (**d**), $R_{2p} = V_0/I$ (**e**), and $\mathcal{V}_H = (V_6 - V_4)/V_0$ (**f**) at $V_{bg} = -0.15$ V and $V_{tg} = 8$ V. **g-i**, Line scans of $\mathcal{V}_{NL}$ along the yellow line in (**a**) upon varying $V_{tg}$ from $-2$ V to 8 V at $V_{bg} = -5$ V (**g**), $V_{bg} = -1$ V (**h**), and $V_{bg} = 3.5$ V (**i**). See Supplementary Video 5 for range of $V_{bg}$ values. **j-l**, Numerical simulations of scanning gate microscopy (see Supplementary Video 9) of $\mathcal{V}_{NL}$ (**j**) and $\mathcal{V}_3$ (**k**) with disordered edge accumulation along the top edge and uniform accumulation along the bottom edge for parameter values $\lambda = 30W$ corresponding to the experimental situation near CNP. See Extended Data Fig. 4 for results away from CNP and for simulations at $\lambda = 8W$ showing large asymmetry between top-right and bottom-right edges. (**l**) Simulation of line scans of $\mathcal{V}_{NL}$ along the top edge upon varying $V_{tg}$. The black curve is an example of edge disorder function $f(x)$ that resembles the experimental situation. It was used to produce the spatially-dependent excess edge line charge density, $f(x)P_e$, along the top edge in **j-l**. See Supplementary Information for parameters.

The suppression of the nonlocality depends on $V_{tg}$, which we examine by measuring $\mathcal{V}_{NL}$ while scanning the SOT repeatedly along the upper edge of the sample (yellow dashed line in Fig. 3a) and varying $V_{tg}$. Figure 3h shows that at each position $x$ along the edge there is a rather sharp $V_{tg}$ threshold (white contour) above which $\mathcal{V}_{NL}$ is strongly suppressed down to vanishingly low values. The threshold $V_{tg}$ shows a rugged structure with sharp variations over as little as 100 nm, apparently limited by the SOT diameter (48 nm) and the scanning height (30 nm). Note that the threshold $V_{tg}$ is always positive, namely the nonlocality is strongly suppressed by depletion of holes along the sample edges, while hole accumulation ($V_{tg} < 0$) has no appreciable effect. A very similar behavior is observed for other negative values of $V_{bg}$ away from CNP where $\mathcal{R}_{NL}$ is large, e.g. $V_{bg} = -5$ V shown in Fig. 3g (see Supplementary Video 5 for a range of $V_{bg}$ values). Surprisingly, when the bulk of graphene is $n$ doped, the behavior remains qualitatively the same as shown in Fig. 3i for $V_{bg} = 3.5$ V corresponding to the peak in $\mathcal{R}_{NL}$ in Fig. 1e: The suppression of the nonlocality still occurs at positive $V_{tg}$ that depletes holes rather than at negative $V_{tg}$ that depletes electrons as one may have expected from symmetry considerations.

While scanning the tip, we also monitor the current $I$ through the sample (Fig. 3d). When the tip is positioned in the interior of the sample, the current remains the same as when the tip is outside the sample, namely the local depletion of holes in the bulk of graphene does not change the sample resistance a lot. However, when the tip is positioned at the edges, the current is drastically suppressed as seen more clearly by plotting $R_{2p} = V_0/I$ in Fig. 3e. Remarkably, by locally depleting the holes at a remote location along the graphene edge, the overall sample resistance increases by over 400 k$\Omega$, tripling its unperturbed value (see Supplementary Video 4 for even larger variations at other $V_{bg}$ values). This finding strongly suggests that a major part of the applied current flows along the graphene edges and that this current channel can be cut off by a local depletion of holes.

We now inspect the individual potentials $\mathcal{V}_3$ and $\mathcal{V}_4$ (Figs. 3b,c). When the edge channel is cut off locally by the tip, the nonlocal transport is discontinued and consequently one may expect the ohmic regime to be restored in which $\mathcal{V}_{NL}$ vanishes and the individual potentials are determined predominantly by the Hall voltage $V_H$ across the sample. As shown by the numerical simulation in Fig. 4c, in the ohmic regime in our configuration the normalized potential $\mathcal{V} = V/V_0$ vanishes in the right part of the sample including $\mathcal{V}_3$ and $\mathcal{V}_4$, while in the left part $\mathcal{V} = 1$ with corresponding normalized Hall voltage $\mathcal{V}_H = V_H/V_0 = 1$. Indeed, when the edge conductance is cut off by the tip along the top edge of the sample, $\mathcal{V}_{NL}$ vanishes (Fig. 3a) and both $\mathcal{V}_3$ and $\mathcal{V}_4$ drop to values close to zero (Figs. 3b,c). Surprisingly, however, when the nonlocal transport is cut off along the bottom edge, the same vanishing of $\mathcal{V}_{NL}$ (Fig. 3a) causes an increase rather than decrease in both $\mathcal{V}_3$ and $\mathcal{V}_4$ to values approaching 1. Therefore, even though the nonlocal transport, as probed by $\mathcal{V}_{NL}$ has been prevented, the ohmic regime has not been restored. Remarkably, in this situation the Hall voltage $\mathcal{V}_H = \mathcal{V}_6 - \mathcal{V}_4$ instead of being restored to 1 may even flip sign and become negative (blue in Fig. 3f). The Hall voltage is usually considered a reliable measure of the sign and the density of the carriers. The observed reversal of the sign of the Hall voltage therefore indicates a highly unconventional transport.

**Discussion of experimental results**

The observations described above, taken together, are hard to reconcile with any of the available models of nonlocal transport. The fact that local depletion of carriers in the bulk of the sample causes no appreciable effect, while depletion along the edge greatly suppresses $\mathcal{R}_{NL}$ is a strong evidence against bulk mechanisms such as anisotropy [39], disorder [25], thermoelectricity [28,29], topological and nontopological bulk spin and valley transport [3,5–9,12,15,17,27,44,45], hydrodynamic flow or ballistic transport [46–48], and electron-hole coexistence [49]. The finding that a local electrostatic perturbation can sharply suppress nonlocal resistance, with the suppression occurring only for one gating polarity, casts doubt on the topologically protected mechanisms due to chiral or helical edge modes [21–23,50]. The observation that the giant nonlocality is not restricted to CNP suggests that the origin of the nonlocality is unrelated to the specific physics of $0^{th}$ Landau level or its magnetism [6,7,12,21–23,26,50,51], or to the presence of both electron and hole carriers [49]. The very small directly imaged local temperature increase and its enhancement upon scanning a depleting tip along the edges renders the thermomagnetic effects [28,29] a highly unlikely source of the giant nonlocality. The clear correlation between the peak of $\mathcal{R}_{NL}$ at CNP and the appearance of dissipation throughout the sample disfavors nondissipative spin or valley currents [1,3,5,12,17,27] or indirect processes involving emission of neutral excitations that are absorbed at the remote contacts [26]. Finally, the finding that the Hall voltage can flip sign due to a local perturbation at a remote location is difficult to explain within the conventional description of the classical or quantum Hall effects.

**Model and simulations of nonlocal transport**

Similarly to band bending and surface states in 3D materials, the states at the edges of atomically thin materials have been discussed extensively [30,33–37]. In graphene, in particular, band bending with $p$-doping of the edges has been suggested to occur due to either intrinsic mechanisms or negatively charged impurities or defects at the edges [34,36,38]. Edge accumulation may also arise due to electrostatic gating by the backgate [35,52], although in this case symmetric electron and hole accumulation should occur for positive and negative $V_{bg}$ with no accumulation at CNP. Here we show that hole accumulation due to intrinsic band bending or negatively charged impurities leads to giant nonlocality and to the intricate transport behavior consistent with the experimental observations.

Without loss of generality, we envision for concreteness a line density of impurities, $N_{imp}$, along the graphene edges, each negatively charged by one electron $e$. They induce a narrow strip of hole accumulation (Fig. 1a) with total line charge density along the edges of $P_e = N_{imp}$ in excess to the uniform bulk hole doping, $p_b = -CV_{bg}$, where $C$ is the backgate capacitance per unit area (see Supplementary Information for discussion of $n$ bulk doping). We further assume for simplicity that the carrier mobility $\mu$ is independent of the carrier density $p$, and hence the local resistivities $\rho_{xx} = 1/pe\mu$ and $\rho_{xy} = B/pe$ are inversely proportional to the local carrier density $p$, and $\rho_{xy}/\rho_{xx} = \mu B \equiv \theta$. We define $\eta$ as the ratio of the total conductance of the two edges and the bulk, $\eta = 2P_e/p_bW = -2P_e/CV_{bg}W$. Solving the current continuity equation and Ohm's law in the nonlocal geometry (Methods), we find that in an infinite strip the solution for $|x| > W$ has a general form $\mathcal{R}_{NL} \cong e^{-|x|/\lambda}$ with the nonlocal decay length

$$\lambda = \frac{W}{\pi}\sqrt{1 + 2\eta\left[1 + \frac{\pi^2\eta}{4}\left(\frac{1+\eta(\theta^2+1)}{\eta+1}\right)^2\right]\bigg/\left[1 + 2\eta\frac{1+\eta(\theta^2+1)}{\eta+1}\right]}. \qquad (1)$$

We inspect the following three regimes: (*i*) In the absence of edge accumulation, $\eta = 0$, the decay length follows the ohmic regime $\lambda = W/\pi$ [1], independent of $B$ and carrier concentration (dashed line in Fig. 1b). (*ii*) In absence of magnetic field, $\lambda = \frac{W}{\pi}\sqrt{1 + 2\eta\left(1 + \frac{\pi^2}{4}\eta\right)/(1 + 2\eta)}$ (cyan curve in Fig. 1b), which reduces for $\eta \ll 1$ to $\lambda = \frac{W}{\pi}(1 + \eta)$ and for $\eta \gg 1$ to $\lambda = \sqrt{\eta}W/2$. This means that at $B = 0$ the nonlocality is enhanced very mildly and even for significant edge accumulation with $\eta = 1$ for which in the local geometry at zero field half of the current flows along the edges, the decay length increases only to $\lambda = 0.58W$ instead of $\lambda = 0.32W$ in the ohmic case. Since $\eta = -2P_e/CV_{bg}W$ decreases with negative $V_{bg}$ due to the increasing bulk conductivity, the nonlocality is largest at CNP and fades with $|V_{bg}|$ (Extended Data Fig. 3). By fitting the zero field data (dashed green line in Fig. 1e), we evaluate $P_e = 1.8\times10^8$ m$^{-1}$, namely an average distance of 5.5 nm between the negatively charged impurities, comparable to previous estimates [43].

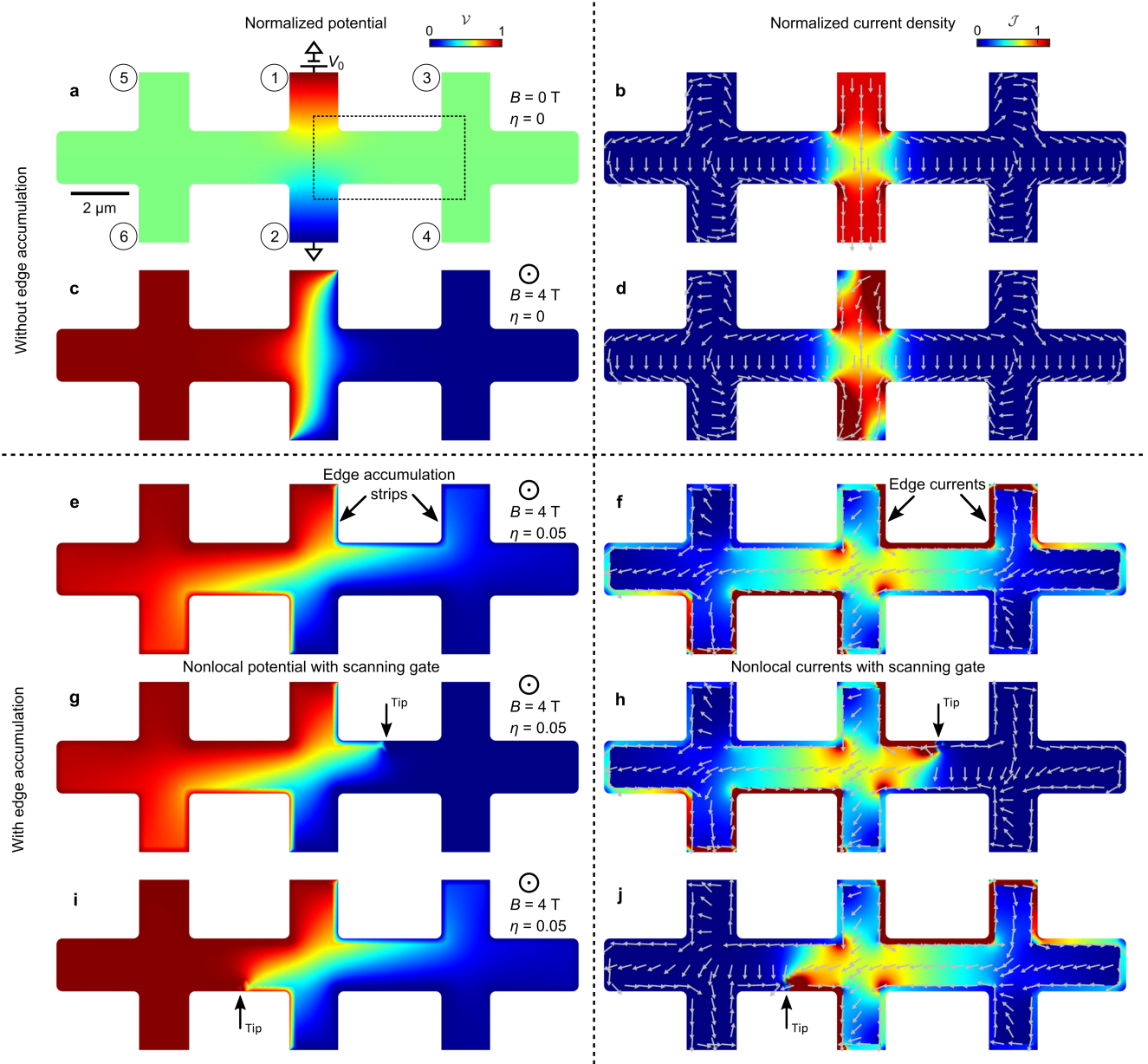

**Fig. 4. Numerical simulations of the nonlocal transport.** Maps of the normalized potential $\mathcal{V} = V/V_0$ (left column) and of the magnitude of the normalized current density $\mathcal{J} = |\mathbf{J}|W/I_0$ (right column) in a $p$-doped sample. The grey arrows of fixed length show the direction of the local current while the magnitude of the current density is color rendered. The width of the edge regions is $w/2 = 200$ nm. In the regions of high current density, $\mathcal{J} > 1.2$, the red color is saturated for clarity. **a-b**, $B = 0$ T in the absence of edge accumulation ($\eta = 0$). The current (**b**) flows in the central vertical strip from source (top) to drain (bottom) down the potential drop (**a**) and decays exponentially into the right and left remote parts of the sample (**b**). The contact numbers are encircled and the dashed rectangle marks the area simulated in Figs. 3j,k in presence of edge accumulation disorder. **c-d**, $B = 4$ T and $\eta = 0$. A transverse Hall potential is established (**c**) which drives the current (**d**) essentially perpendicular to the potential gradient with spatial distribution equivalent to that in **b** (except near the contacts). **e-f**, $B = 4$ T and $\eta = 0.05$, $\lambda = 2.54W$. (see Supplementary Videos 6 and 7 for range of $\eta$ and $B$ values). The current (**f**) flows out of the source along top-right edge (red) and gradually leaks into the bulk where it reverses its direction and flows to the left with $x$ component of the current flowing against the potential drop (**e**). The current is then gradually absorbed into the bottom-left edge where it reverses its direction again and flows to the right towards the drain. **g-h**, $B = 4$ T and $\lambda = 2.54W$ with a depleting tip at the location marked by the arrow at the top-right edge. The flow of the current (**h**) from the source along the top-right edge (red) is terminated at the tip location and is diverted into the bulk with enhanced local current density. The nonlocal potential profile (**g**) becomes flat to the right of the tip (blue). **i-j**, Same as **g-h** with the tip at the bottom-left edge. The depleting tip prevents the current (**j**) from accumulating along the entire left edge between the source and the tip location. The nonlocal potential (**i**) flattens out to the left of the tip (red). See simulations of a scanning tip in Supplementary Videos 8 and 9.

The most interesting regime (*iii*), however, arises at finite $B$ and finite edge accumulation, where in the limit of $\eta\sqrt{\frac{\theta^2+1}{\eta+1}} \gg 1$ we find $\mathcal{R}_{NL} = e^{-|x|/\lambda}$ with $\lambda = \frac{\eta W}{2}\sqrt{\frac{\theta^2+1}{\eta+1}}$, which leads to giant nonlocality (Fig. 1b). At $B = 5$ T and our carrier mobility $\mu \cong 2.6\times10^5$ cm$^2$V$^{-1}$s$^{-1}$ we arrive at $\theta = \mu B \cong 130$. This implies that near CNP, where we find $\eta \cong 0.7$ (Extended Data Fig. 3a), for which at $B = 0$ the nonlocality increases only mildly, at 5 T a giant nonlocality takes place with decay length $\lambda \cong 35W$ greatly exceeding the sample size. Even at $V_{bg} = -10$ V, at which $\eta = -2P_e/CV_{bg}W \cong 0.02$, the nonlocality remains very large with $\lambda \cong 1.29W$ so that $\mathcal{R}_{NL}$ exceeds the ohmic value by three orders of magnitude (Fig. 1e and Extended Data Figs. 3b,c).

Analytic calculations (Methods) and numerical COMSOL simulations (Fig. 4) provide detailed insight into the nonlocal transport mechanism driven by a magnetic-field-induced bulk-edge decoupling. In the absence of edge accumulation (ohmic regime, $\eta = 0$) at $B = 0$ the potential $\mathcal{V}$ in the remote right and left arms is essentially constant (Fig. 4a) and the current density ***J*** flows mainly along the central vertical strip following the potential drop (Fig. 4b). At large $B$, the Hall voltage and the corresponding transverse electric field become dominant (Fig. 4c), but the current distribution in the bulk is unchanged (except for the focusing at the hotspots at the source and drain contacts) with essentially unaltered ohmic value of $\mathcal{R}_{NL}$ (Fig. 4d and Supplementary Video 6). With edge accumulation ($\eta > 0$), in contrast, the current flow becomes very intriguing. In contrast to the common expectation in the context of the QH effect, in which the current flows downstream along the high-potential edge of the sample (top-left), the current emerges out of the source (top contact) predominantly along the top-right edge (Fig. 4f), and then flows in the $\hat{x}$ direction along the top edge of the right arm. Instead of continuing its flow along the higher conductivity edge, the current gradually leaks out of the edge into the bulk. However, rather than flowing in the $-\hat{y}$ direction towards the bottom edge and then to the drain as is the case at $\eta = 0$ (Figs. 4b,d), most of the bulk current flows in the $-\hat{x}$ direction against the flow in the top edge (Fig. 4f and Extended Data Fig. 2). Remarkably, upon reaching the central vertical strip of the sample the current does not flow to the drain. Instead, the $x$ component of the current flows "upstream", against the potential drop, into the left arm where it is gradually absorbed by the edges. The current then reverts its flow again to the $\hat{x}$ direction and reaches the drain through the bottom-left edge (see Extended Data Fig. 2 and Supplementary Videos 6 and 7 for more details). These long-range nontopological currents give rise to a giant nonlocal resistance and to dissipation proliferating over the entire sample as observed experimentally in Fig. 2.

This unusual nonlocal transport is very sensitive to disorder and to local perturbations. At $B = 0$, if the edge channel is disrupted locally by disorder or by the tip potential, the current tends to bypass the disruption through the bulk and return to the edge. In presence of elevated magnetic field, however, the rules change completely resulting in a unidirectional bulk-edge decoupling (Supplementary Information): In the right half of the sample a current that leaves the edge channel is not allowed to return to it. The current then flows along the edge from the source up to the interruption point where it is diverted into the bulk, while the rest of the channel carries no current. Such interruption of the edge current by the depleting tip causes a sharp drop in $\mathcal{R}_{NL}$ as observed experimentally in Fig. 3 and described by simulations in Figs. 4g,h and in Supplementary Videos 8 and 9. Remarkably, an opposite rule applies to the left half of the sample – the edge can only gutter the current from the bulk and any current that has been absorbed by the edge channel cannot be released. Therefore, an interruption along the channel prevents current from being adsorbed by the channel over its entire length from the source up to the interruption point, and only the remaining part of the channel gathers the current from the bulk into the drain (Fig. 4j). These rules have major global effects as observed in Fig. 3 and simulated numerically in Figs. 3j-l, including the tip-induced increase in the two-probe resistance, high sensitivity to edge disorder, and inversion of the polarity of the Hall voltage (see Methods and Extended Data Fig. 5).

**Implications**

In contrast to conventional gapped semiconducting 2D electronic systems in which band bending can cause either edge charge depletion as in GaAs heterostructures or edge accumulation as in InAs system [53], in Dirac materials any type of band bending causes edge accumulation near CNP. The revealed nonlocal transport mechanism due to magnetic-field-induced bulk-edge decoupling in presence of edge accumulation is completely generic, and thus applicable to a wide range of van der Waals 2D structures and may extend to room temperature in high mobility devices [1]. The nonlocality grows with the edge-to-bulk conductivity ratio $\eta$, that diverges when the bulk carrier concentration vanishes at CNP [1–4,6,7,21–24], or upon opening a bulk energy gap in moiré superlattices [14,16–18,30] or in bilayer graphene in presence of a displacement field [12,15,27], or upon approaching conduction band edge in transition-metal dichalcogenides [9], possibly leading to large nonlocal effects even at zero magnetic field which then are greatly enhanced at finite $B$. Being very sensitive to edge disorder and the amount of edge accumulation, the edge current mechanism can also explain large sample-to-sample and contact-to-contact variability [1,24,29], as well as a peculiar dependence on the magnetic field direction [1] (Supplementary Information). We envision that spatial confinement of edge accumulation may lead to 1D-like transport including resistance quantization [21,23,50]. Finally, the edge accumulation may contribute to the edge-confined supercurrents observed in graphene proximity devices [34,36,40] or even provide an alternative mechanism if edge doping is sufficiently high. The surprising observation of long-range edge currents not protected by topology, but nevertheless robust and coexisting with the bulk conduction, calls for careful re-examination of the numerous reported nonlocal transport phenomena, offering a possible simpler explanation to be considered in future experiments on transport in quantum materials.

**Methods**

**Device fabrication**

The graphene heterostructure was fabricated using a modified dry flip-stack method. After exfoliating natural graphite and hBN flakes onto oxidized silicon wafer (290 nm of SiO$_2$), thick hBN is picked up using a PDMS/PPC stamp at 50 degrees. Graphene and a thin (typically <10 nm) hBN layer are then picked up successively to form an hBN/Gr/hBN stack. Next, instead of directly releasing the stack onto either SiO$_2$ surface or pre-exfoliated graphite on SiO$_2$, at ~65-75°C the stack is instead released onto a second PDMS/PMMA stamp. At 65-75°C the adhesive force between the PPC and hBN/Gr/hBN stack is significantly weaker than the PMMA adhesive force, therefore the stack can easily detach onto the second stamp. Finally, the stack is released onto a final graphite flake on SiO$_2$ at 145°C, which acts as a backgate separated from graphene by a thick hBN. The advantage of this flip-stack method over previous methods is fourfold: (i) The process is compatible with van der Waals heterostructures wherein the capping (top) layer needs to be extremely thin – even monolayer. In traditional methods, dry stamping monolayer capping layers poses a significant challenge due to strain-induced crack formation, whereas in this modified method macroscopic forces are stabilized by the initial thick hBN. (ii) Yield for stamping of thicker flakes is far higher than for extremely thin flakes, improving overall sample production. (iii) The final transfer can be done at up to 250°C since the stamp is composed of PMMA and not PPC or PC, and therefore interlayer contamination is greatly reduced. (iv) During the release step, the PMMA is peeled off of the top hBN layer, leaving a significantly cleaner surface than for samples where PC or PPC is released onto the substrate and removed via solvent.

For device fabrication, electron beam lithography (Vistec 5200+) was used to define both the contact location and sample geometry, followed by CHF$_3$ and O$_2$ reactive ion etching. Contacts were defined by a predominately physical etching (20 W RIE) with ~0.75 nm/s etch rate allowing controlled etch of contacts without reaching the underlying graphite gate. Mesa etching incorporated a chemical/physical etch (5W RIE, 150 W ICP) followed by a short physical RIE process (20 W RIE, 0 W ICP). All contact metallization was done via sputtering of Cr/Au (1 nm/60 nm) and standard liftoff procedure. After the final fabrication steps, the sample was soaked in a tetramethylammonium hydroxide (TMAH)-based alkaline developer (MIF-319) and AFM-cleaned to remove residual surface contamination from the fabrication process.

The sample was patterned into a form of a Hall bar (optical image in Fig. 1a inset) with the central horizontal bar of width $W \cong 1.8$ μm, length of 17.7 μm, and a distance between the centers of the contacts 1-2 to 3-4 of ~5.1 μm.

**SQUID-on-tip, thermal imaging, and scanning gate microscopy**

The SOTs were fabricated using three-step deposition of superconducting film on a pulled quartz pipette as described previously [54,55]. A Pb SOT of 48 nm diameter, fabricated using thermal deposition [56], displayed thermal sensitivity of 1.9 μK/Hz$^{1/2}$ and operated in field of up to 1 T, while MoRe SOT of 100 nm diameter, fabricated using a recently introduced method of collimated sputtering [57], had thermal sensitivity of 1.1 μK/Hz$^{1/2}$ at 0 T and operated in field of up to 5 T with sensitivity of 3.3 μK/Hz$^{1/2}$. The thermal imaging (as well as the rest of the measurements) was carried out at 4.2 K in presence of about 60 mbar pressure of He exchange gas as described previously [32,38,43]. A sinusoidal *ac* current $I_0$ in the range of 10 nA to 10 μA was applied to the sample at frequency $f = 66.66$ Hz and the corresponding thermal map $T_{2f}(\boldsymbol{r})$ of the sample is acquired by lock-in amplifier locked to the second harmonic frequency $2f$. As a result, the $T_{2f}(\boldsymbol{r})$ image provides a map of the current-induced local temperature increase in the sample. In order to compare the thermal images at various fields and $V_{bg}$ for which $R_{2p}$ varies by several orders of magnitude, the *ac* voltage $V_0$ applied to the sample was adjusted for each image to keep the same total dissipated power $P = V_0 I_0 = 15$ nW. The thermal imaging was acquired at a constant scanning height of 130 to 160 nm above the surface of the top hBN with typical pixel size of 46 nm, acquisition time of 40 ms/pixel, and image size of 85 × 376 pixels/image.

For scanning gate imaging [33,37,38,58–60] a *dc* voltage $V_0$ of about 4.5 to 5.5 mV chopped by a square wave at a frequency of 25 Hz was applied to the sample and the nonlocal differential potential $V_{NL}$, the current $I$ through the sample, and the individual potentials $V_i$ at the different contacts $i$ were measured using lock-in amplifiers vs. SOT position, $V_{bg}$, and $V_{tg}$. For a meaningful analysis of nonlocal transport, the measured potentials should be normalized by the applied $V_0$. For a higher accuracy analysis, we have corrected the actual $V_0$ applied to graphene by subtracting from the applied voltage the potential drop on the contact resistances, $2R_c I$. In most of the cases $R_c \ll R_{2p}$ and hence this correction is completely negligible except for the highest values of $|V_{bg}|$ at zero field, where the sample resistance drops to few hundred Ohms. The scans in Fig. 3 were acquired at a constant scanning height of 80 to 110 nm above the surface of the top hBN with typical pixel size of 32 nm, acquisition time of 40 ms/pixel, and image size of 110 × 319 pixels/image.

The piezoelectric scanners (attocube) have a slight hysteresis and corresponding distortions which depend on the scanning range and speed. The 2D images in Figs. 2 and 3 were acquired with fast scanning axis along the $y$ direction, while for line scans along the graphene edges in Figs. 3g-i the fast scanning axis was along $x$ direction. The $x$ scale in Figs. 3g-i was accordingly corrected for the distortions to match the 2D images.

**Nonlocal transport measurements**

Two-probe and four-probe transport measurements were performed using standard lock-in techniques using either *ac* current at 7.57 Hz or *dc* current chopped by a square wave at 25 Hz. In addition, single-ended potentials at the different contacts were measured with respect to ground in the scanning gate imaging. From Hall and longitudinal measurements we derive the backgate capacitance $C = 1.5\times10^{-8}$ F/cm$^2$ = $9.4\times10^{10}$ e/cm$^2$V and carrier mobility $\mu = 2.6\times10^5$ cm$^2$/Vs with weak dependence on carrier concentration. By comparing the potentials at the different leads we estimated the contact resistance $R_c$ to be about 80 Ω when graphene is *n*-doped and 180 Ω for *p*-doping. By varying the applied current, we have confirmed that for most of our results the nonlinear effects were small, although the revealed mechanism of conductance along the edges in presence of edge disorder can readily cause significant nonlinearities.

Extended Data Figs. 1a,b present $R_{NL}$, $R_{2p}$, and $\mathcal{R}_{NL}$ at $B = 3$ and 5 T for negative $V_{bg}$. $R_{NL}$ shows a rapid decrease with $|V_{bg}|$ and pronounced dips at QH plateaus where $R_{2p}$ is quantized. The rapid decrease of $R_{NL}$, however, is governed to a large extent by the decrease in $\rho_{xx} = 1/pe\mu \propto 1/|V_{bg}|$, which can be partially corrected for by normalizing by $R_{2p}$. The resulting $\mathcal{R}_{NL} = R_{NL}/R_{2p}$ shows a moderate decrease with $|V_{bg}|$ revealing that the nonlocality is not restricted to the vicinity of CNP but is rather a broad phenomenon. Moreover, the periodic dips in $\mathcal{R}_{NL}$ do not indicate vanishing of the edge accumulation at specific values of $V_{bg}$. This is demonstrated in Extended Data Fig. 1c showing $\mathcal{R}_{NL}$ at $B = 1$, 1.2, and 1.4 T. By normalizing the envelopes of 1 and 1.2 T data (dotted curves) to that of 1.4 T, Extended Data Fig. 1d shows that the nonlocal mechanism is continuous as described by the dotted envelope curve and decreases monotonically with $|V_{bg}|$ (and increases monotonically with $B$) and its observation is merely masked periodically in $V_{bg}$ due to vanishing of $\rho_{xx}$ at the QH plateaus while $R_{2p}$ remains finite.

**Analytic solution of nonlocal transport with edge accumulation**

In a uniform sample the continuity equation $\nabla \cdot \mathbf{J} = 0$ and Ohm's law $\mathbf{J} = \sigma \mathbf{E} = -\sigma \nabla V$ yield the Laplace equation $\nabla^2 V = 0$, which is readily solvable (here $\mathbf{E}$ is the electric field and $\sigma$ is the conductivity tensor given by $\sigma_{xx} = \rho_{xx}/(\rho_{xx}^2 + \rho_{xy}^2)$ and $\sigma_{yx} = \rho_{xy}/(\rho_{xx}^2 + \rho_{xy}^2) = \theta \sigma_{xx}$. For a sample of width $W$ in the $y$ direction and infinitely long in $x$ with current injection points at $(0, \pm W/2)$, the solution (see Supplementary Information) is $R_{NL}(x) = \frac{\rho_{xx}}{\pi} \ln\left[\frac{\cosh(\pi x/W)+1}{\cosh(\pi x/W)-1}\right]$, which reduces to $R_{NL}(x) = \frac{4\rho_{xx}}{\pi} e^{-\pi|x|/W}$ for $|x| \gg W$ [1].

In presence of charge accumulation, however, the conductivity tensor $\sigma$ is position dependent and hence the Laplace equation is no longer applicable, rendering the general solution difficult. To overcome this limitation, we approximate the edge accumulation by strips of width $w/2$ along the edges with uniform carrier concentration $p_e$ which is higher than $p_b$ in the bulk (Extended Data Fig. 2a). As a result, the Laplace equation holds separately in the edge and bulk regions and thus can be solved using appropriate boundary conditions for $\mathbf{J}$ and $\mathbf{E}$ at the interfaces. Taking the limit of $w \to 0$ while keeping the excess line charge in the edge regions $P_e$ constant, in the limit of $\eta\sqrt{\frac{\theta^2+1}{\eta+1}} \gg 1$ we arrive at the following solutions for the potentials $V_{top}(x), V_{bot}(x)$ and the currents $I_{top}(x), I_{bot}(x)$ along the top and bottom edges, as well as for the total current $I_{bulk}(x)$ flowing in the $\hat{x}$ direction in the bulk (see derivation in Supplementary Information):

$$\lambda = \frac{\eta W}{2} \sqrt{\frac{\theta^2+1}{\eta+1}}$$

$$\mathcal{V}_{top/bot}(x) = \frac{V_{top/bot}(x)}{V_0} = \frac{1}{2} \pm \frac{1}{2}\left(1 + \text{sgn}\, x \frac{\theta}{\sqrt{(\eta+1)(\theta^2+1)}}\right)(e^{-|x|/\lambda} - 1) \pm \frac{1}{2}$$

$$\mathcal{I}_{top/bot}(x) = \frac{I_{top/bot}(x)}{I_0} = \frac{1}{2}\left(\pm \text{sgn}\, x + \frac{\theta}{\sqrt{(\eta+1)(\theta^2+1)}}\right) e^{-|x|/\lambda}$$

$$\mathcal{I}_{bulk}(x) = \frac{I_{bulk}(x)}{I_0} = -\frac{\theta}{\sqrt{(\eta+1)(\theta^2+1)}} e^{-|x|/\lambda}$$

$$\mathcal{R}_{NL}(x) = \frac{R_{NL}(x)}{R_{2p}} = \frac{V_{NL}(x)}{V_0} = \mathcal{V}_{NL} = \mathcal{V}_{top}(x) - \mathcal{V}_{bot}(x) = e^{-|x|/\lambda}$$

$$R_{2p} = \frac{V_0}{I_0} = \rho_{xx} \sqrt{\frac{\theta^2+1}{\eta+1}}$$

These expressions can be also derived using a simpler model of coupled 1D conduction channels (Supplementary Information).

Extended Data Fig. 2c shows the potentials along the edges $\mathcal{V}_{top}$ and $\mathcal{V}_{bot}$ for the case of $\eta = 0.2$ and $\theta = 26$ for which $\lambda = 2.4W$. The potentials decay exponentially reaching values determined by the Hall voltage $\mathcal{V}_H = \mathcal{V}_{top}(x=-\infty) - \mathcal{V}_{top}(x=\infty) = 1/\sqrt{\eta+1}$, which is suppressed relative to $\mathcal{V}_H = 1$ without edge accumulation. Note that in samples of finite length the suppression of the Hall voltage will be larger and in presence of edge disorder $\mathcal{V}_H$ can even flip its sign as observed experimentally and described below.

For small $\eta$ and $\theta$ the condition $\eta\sqrt{\frac{\theta^2+1}{\eta+1}} \gg 1$ is violated. For this case, we derive in Supplementary Information an improved approximation for $\lambda$ given by Eq. (1) that is valid for any value of $\eta$ and $\theta$.

Unprotected edge states should decay exponentially due to localization or coupling to the bulk. The above results are consistent with this fundamental notion. In zero magnetic field, the decay length $\lambda$ is very short due to strong bulk-edge coupling. The above derivation shows, however, that the magnetic field gives rise to field-induced unidirectional bulk-edge decoupling mechanism (see Supplementary Information for discussion) leading to the linear increase of $\lambda$ with $\theta$ and hence exponential growth of $R_{NL}$ with magnetic field in an infinite sample. In finite samples the exponential growth of $R_{NL}$ with field will saturate once $\lambda$ becomes larger than the sample size.

**Description of the nonlocal current flow in the presence of edge accumulation**

The above derivation reveals intriguing current flow behavior as shown in Extended Data Figs. 2b,d. Naively, in the QH effect the current is expected to flow from source (top) to drain (bottom) along the high-potential edge (left) as dictated by the downstream chiral direction. With edge accumulation, however, the opposite situation takes place in which most of the current flows from the source along the upstream chiral direction (right). The ratio of the right-flowing and left-flowing currents is $\frac{\mathcal{I}_{top}(x=0^+)}{|\mathcal{I}_{top}(x=0^-)|} = \frac{(1+\sqrt{\eta+1})^2}{\eta} \cong 4/\eta$, which means that for small $\eta$ essentially all the current exits the source contact along the upstream edge. In the case of $\eta = 0.2$, presented in Extended Data Fig. 2d, 96% of $\mathcal{I}_{top}$ emerges to the right (red curve at $x > 0$) and 4% to the left (red curve at $x < 0$). The edge current $\mathcal{I}_{top}(x)$ decays exponentially due to leakage into the bulk. One would then expect the current to flow through the bulk to the bottom edge and then to the drain, as is the case in zero magnetic field. Instead, the current that leaks into the bulk reverses its direction and flows as $\mathcal{I}_{bulk}$ predominantly in $-\hat{x}$ direction (green curve), while only a small fraction of it is collected in the bottom edge, $\frac{\mathcal{I}_{bot}(x>0)}{\mathcal{I}_{bulk}(x>0)} = \frac{1}{2}(\sqrt{\eta+1} - 1) \cong \eta/4$ for small $\eta$ (blue curve at $x > 0$). Moreover, upon reaching the central part of the sample, most of the bulk current $\mathcal{I}_{bulk}$ does not flow to the drain, but rather continues uninterrupted into the left part of the sample (green curve) despite the fact that the potential there is higher (Extended Data Fig. 2c). So the edge currents flow with the direction of the electric field $E_x = -\partial_x V$, given by $I_{top}(x) = -\partial_x V_{top}/R_{edge}$ and $I_{bot}(x) = -\partial_x V_{bot}/R_{edge}$ (red and blue arrows in Extended Data Fig. 2c), where $R_{edge} \cong 1/(P_e e \mu)$ is the edge resistance per unit length, while the $x$ component of the bulk current $I_{bulk,\hat{x}}(x)$ flows against $E_x$ (green arrow in Extended Data Fig. 2c).

One can understand this behavior as following. In a conventional Hall effect the transverse boundary conditions are equivalent to an "open circuit". As a result, a Hall potential is formed which counterbalances the Lorentz force resulting in a net zero transverse current. The Lorenz force acts as a generator that drives the carriers from the low-potential edge to the high-potential edge against the Hall potential drop, while the Hall component of the electric field inhibits this transverse current, which would otherwise flow in a "closed circuit" case. In the presence of edge accumulation, the edges act as 1D conductive channels that can absorb or supply current to the bulk. As a result, the transverse boundary conditions of the bulk are no longer of an "open circuit", thus a transverse current can now flow from one edge to the other, resembling a partially "closed circuit". Being driven by the Lorentz force, the transverse current in the bulk can only flow from the low-potential edge to the high-potential edge. In the nonlocal configuration in an infinite sample, in order to take advantage of the enhanced conductivity of the edges, the current thus has to exit the source along the low-potential edge, traverse the bulk against the Hall potential drop, and then flow to the drain along the high-potential edge. The directionality of the Lorentz force also dictates that a current that has left the 1D low-potential edge due to tip depletion or edge disorder cannot return to it and the current that has entered the 1D high-potential edge cannot leave it. Note that the transverse current $J_x$ that flows against $E_x$ is accompanied by the longitudinal current $J_y$ that flows along $E_y$ such that the transport remains always dissipative with $\boldsymbol{E} \cdot \boldsymbol{J} > 0$.

These complex current patterns are closely related to the previous studies [61–63] that have shown that in inhomogeneous samples at large $\theta$ the current tends to focus into regions of highest $\sigma_{xy}$. Such inhomogeneities with enhanced $\sigma_{xy}$, however, have been previously considered to occur in the bulk whereas in our case they are present along the edges. Note that the derived solutions are based on classical transport description without any topological or QHE considerations.

**Description of the edge charge accumulation**

Edge states and band bending in 2D materials have been discussed extensively [30,33–36,64–67]. Edge charge accumulation may arise due to electrostatic gating by the backgate [35,52,68–71], in which case there is no accumulation at CNP and symmetric electron and hole accumulation for positive and negative $V_{bg}$, or due to intrinsic edge states or extrinsic band bending [30,33,34,36–38,40,64,65,72–87], which may give rise to charge accumulation at CNP and asymmetric behavior between $p$ and $n$ doping. Our results are consistent with the latter, although at large $|V_{bg}|$ electrostatic gating may become important (see Supplementary Information).

We describe the edge accumulation by an excess hole line charge $P_e$ induced by negatively charged defects along the graphene edges. In the limit of $w \to 0$, the edge to bulk conductance ratio is given by $\eta(V_{bg}) = 2\mu_{edge}P_e/\mu W p_b(V_{bg})$, where $\mu$ is the bulk carrier mobility and $\mu_{edge}$ is the mobility of the carriers in the edge regions that is generally expected to be lower than $\mu$ due to enhanced edge scattering. We assume for simplicity $\mu_{edge} = \mu$. This assumption, however, does not alter the derived analytic expressions since in the $w = 0$ limit they depend only on $\eta$ and on the bulk $\theta$ (see Supplementary Information)

$\eta(V_{bg})$ should diverge at CNP due to vanishing of $p_b(V_{bg})$ at $V_{bg} = 0$. It is commonly believed, however, that at CNP the carrier concentration remains finite due to bulk charge disorder $p_{dis}$. We therefore describe the bulk carrier concentration for $p$-doping as $p_b(V_{bg}) = \sqrt{p_{dis}^2 + (CV_{bg})^2}$, where $C = 9.4 \times 10^{10}$ e/cm²V is our device backgate capacitance. By fitting to the experimental $\mathcal{R}_{NL}$ data at $B = 0$ T (green dashed line in Extended Data Fig. 3c), we evaluate $P_e \cong 1.8 \times 10^8$ m$^{-1}$ and $p_{dis} \cong 2.9 \times 10^{10}$ cm$^{-2}$. Extended Data Fig. 3a shows the resulting $\eta(V_{bg})$, which at CNP is limited by bulk charge disorder, $\eta(0) = 2P_e/p_{dis}W \cong 0.7$ and drops with negative $V_{bg}$ to 0.02 at $V_{bg} = -10$ V.

Using the derived $\eta(V_{bg})$, we present in Extended Data Fig. 3b the corresponding nonlocal decay length $\lambda(V_{bg})$ at various $B$, calculated from Eq. (1). At zero field $\lambda = \frac{W}{\pi}\sqrt{1 + 2\eta\left(1 + \frac{\pi^2}{4}\eta\right)/(1 + 2\eta)}$ is increased only mildly above the ohmic value (cyan curve), however, with increasing $B$, $\lambda \cong \frac{\eta W}{2}\sqrt{\frac{\theta^2+1}{\eta+1}}$ grows sharply due to the $\theta = \mu B$ term, which reaches a value of 130 at $B = 5$ T with our carrier mobility $\mu \cong 2.6 \times 10^5$ cm²V$^{-1}$s$^{-1}$. The resulting $\lambda$ (red curve) readily becomes comparable to distance to the nonlocal contacts (2.85$W$ in our case marked by the dotted line in Extended Data Fig. 3a) and to the sample size and can even greatly exceed it, leading to giant nonlocality. The dashed curves in Extended Data Fig. 3c show the corresponding calculated $\mathcal{R}_{NL}$, which show good qualitative agreement with the experimental data. The experimentally evaluated $P_e$ and the corresponding $\eta \cong 0.7$ at CNP should be considered as lower bounds since $\mathcal{R}_{NL}$ is governed by points of weakest charge accumulation along the edges. Note that edge accumulation that is even two orders of magnitude lower than the value estimated in our sample will still result in giant nonlocality at CNP at fields above 5 T as shown in Extended Data Fig. 3d. Also cleaner devices with lower bulk $p_{dis}$ will give rise to larger $\eta$ at CNP.

**Numerical simulations of nonlocal transport in finite sample**

Numerical simulations of a finite sample with geometry equivalent to the experimental device were carried out using COMSOL as described in the Supplementary Information. These results corroborate the analytic expressions while revealing differences between finite and infinite samples and allow exploration of the effects of edge disorder. Supplementary Video 6 compares the transport behaviors with and without edge accumulation. For $\eta = 0$, upon increasing $B$, the electric field ($\boldsymbol{E} = -\boldsymbol{\nabla}V$) rotates from $-\hat{y}$ to $\hat{x}$ directions as dictated by the Hall angle, however, the current distribution in the bulk is directed predominantly along $-\hat{y}$ independent of $B$, and the transport remains ohmic. For $\eta = 0.1$ and low $B$ the transport is very similar to the ohmic case. Upon increasing $B$, however, the behavior changes profoundly. The electric field and current directions in the bulk rotate and the current flows along the top-right edge, leaks into the bulk, traverses to the left side of the sample against the potential rise, drains into the bottom-left edge, and flows towards the drain leading to the giant nonlocality.

In an infinite sample the roles of the edges are well defined: the top edge acts as an extended current source that seeps the current into the bulk while the bottom edge acts as a gutter that gradually drains the current from the bulk and channels it towards the drain. In a finite size sample, however, the top and bottom edges are connected and thus their functions become more intricate. This is exemplified more clearly in Supplementary Video 7 for $B = 2$ T vs. $V_{bg}$ or $\lambda$. For small $\lambda$ ($V_{bg} = -10$ V) the transport is similar to the ohmic regime (Figs. 4c,d) with electric field along $\hat{x}$ and current along $-\hat{y}$. Upon increasing $\lambda$, the bulk current rotates it direction towards $-\hat{x}$ and reaches its maximal value (orange) when $\lambda$ becomes comparable to the sample size ($V_{bg} \cong -1$ V). In contrast to infinite sample, however, this picture collapses sharply when $\lambda$ grows further ($V_{bg} = 0$ V): the bulk current vanishes and almost the entire current is carried by the two edges all the way from the source to the drain. Remarkably, in this situation the electric field in the bulk is rotated by 90° relative to the ohmic regime in Fig. 4c. Moreover, the potential profile along $y$ in the right and left arms shows a nonmonotonic behavior reflecting the fact that the current in the bulk flows in direction that is opposite to the current flow in the top-right and bottom-left edges.

**Inhomogeneous edge charge accumulation**

Figures 3a-i show that the edge accumulation is highly disordered. To describe an inhomogeneous edge charge accumulation we define the edge disorder function $f(x)$ that determines the spatially-dependent excess edge line charge density, $f(x)P_e$, where $0 \leq f(x) \leq 1$. In the case of uniform edge accumulation, $f(x) = 1$, the tip potential suppresses the edge current for any value of positive $V_{tg}$. In presence of edge disorder, however, the points of weakest edge accumulation form bottlenecks for the edge current thus limiting the current that flows along the rest of the edge. In this situation, the effect of $V_{tg}$ becomes position dependent as observed in Figs. 3g-I and Supplementary Video 9. At points of weakest edge accumulation (see $f(x)$ depicted in Fig. 3l), positive $V_{tg}$ suppresses the current and the measured $\mathcal{V}_{NL}$. At points of large edge accumulation, however, the edge current is not limited by the value of the local edge accumulation and hence suppression by $V_{tg}$ does not occur until a threshold is reached beyond which the tip-induced depletion creates a new bottleneck for the edge current, as described by the simulations in Fig. 3l and in accord with the experimental results in Figs. 3g-i.

The depleting tip diverts the edge current into the bulk as seen in Supplementary Video 8, creating a hotspot of higher bulk current density and larger potential gradient, which causes the observed enhancement of the thermal signal along the edges in Figs. 2e,g and in Supplementary Video 3 upon application of a positive $V_{tg}$. The hotspot is enhanced at points of weaker edge accumulation as shown by the numerical simulations in Supplementary Video 9, giving rise to the inhomogeneous thermal signal along the edges in Figs. 2e,g and in Supplementary Video 3.

Supplementary Video 4 shows that $R_{2p}$ can reach values in excess of 1 MΩ in the vicinity of CNP. Such a high resistance is often taken as an evidence for the presence of an energy gap throughout the sample including the edges. Our results show that in high-mobility samples in presence of edge accumulation such resistance values are readily achievable upon increasing the magnetic field without gap opening.

**Methods References**

**Acknowledgments** The authors thank M. E. Huber for SOT readout system, S. Grover for data acquisition setup, and G. Zhang, I. V. Gornyi, Y. Gefen, and A. Uri for discussions. This work was supported by the European Research Council (ERC) under the European Union's Horizon 2020 research and innovation program (grants No 785971 and 786532), by the Israel Science Foundation ISF (grants No 921/18 and 994/19), by the Sagol Weizmann-MIT Bridge Program, by the German-Israeli Foundation for Scientific Research and Development (GIF) Grant no. I-1505-303.10/2019, and by Lloyd's Register Foundation. EZ acknowledges the support of the Andre Deloro Prize for Scientific Research and Leona M. and Harry B. Helmsley Charitable Trust grant 2018PG-ISL006.

**Author contributions** A.A., A.M., and E.Z. conceived the experiments. D.J.P. and A.K.G. provided the studied devices. A.M. and A.A. carried out the measurements and data analysis. K.B. and Y.M. fabricated the SOTs and the tuning fork feedback. T.H., A.A., and L.S.L. developed the analytic models. A.A. performed the numerical simulations. A.A., A.M., T.H., E.Z., L.S.L, and A.K.G. wrote the manuscript with contributions from the rest of the authors.

**Data availability** The data that support the findings of this study are available from the corresponding authors on reasonable request.

**Competing interests** The authors declare no competing interests.

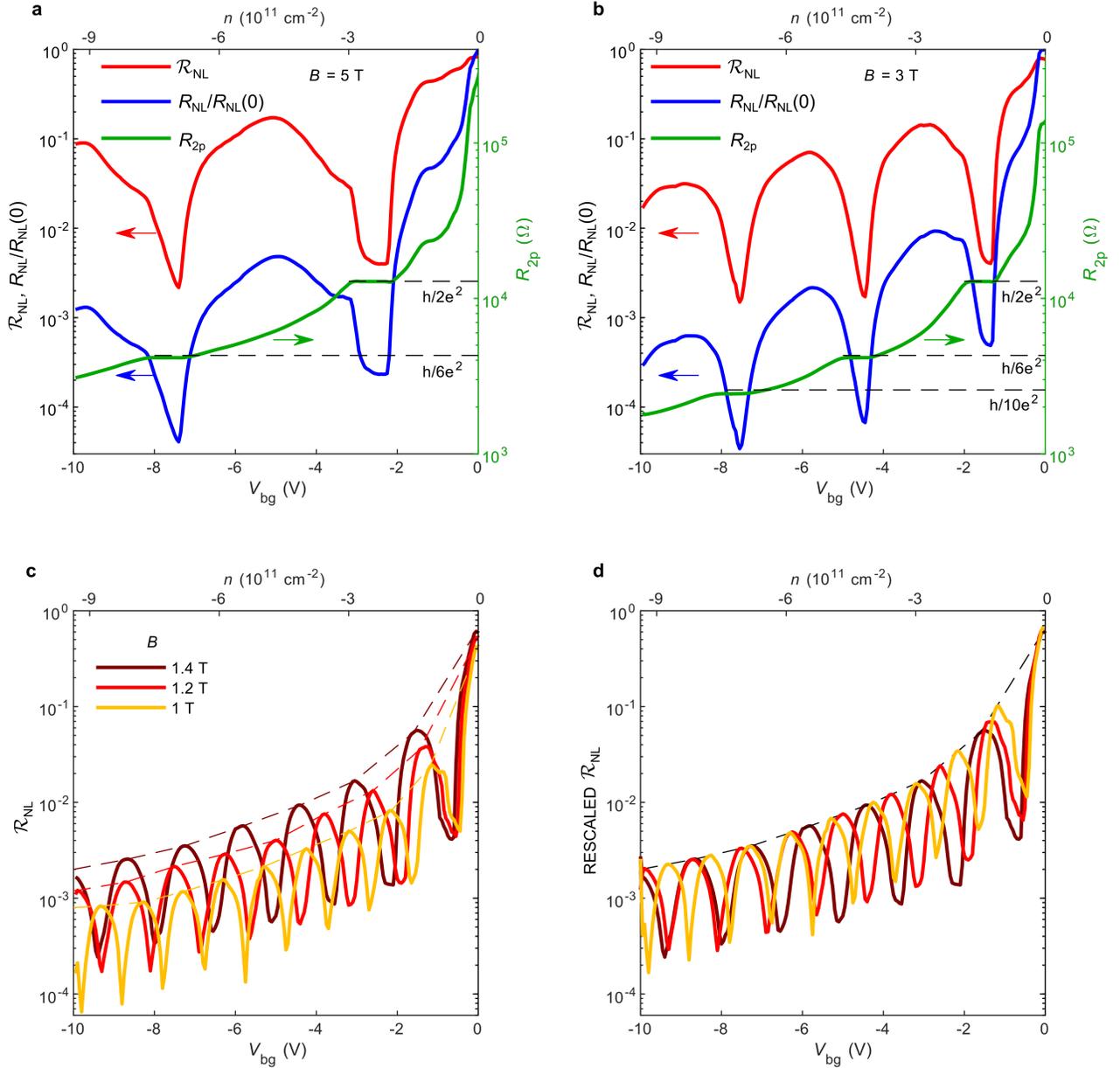

**Extended Data Fig. 1**. **Nonlocal transport measurements. a-b**, The two-probe resistance $R_{2p}$ (right axis) and the nonlocal resistances $R_{NL}$ and $\mathcal{R}_{NL} = R_{NL}/R_{2p}$ (left axis) at $B = 5$ T (**a**) and 3 T (**b**) for negative $V_{bg}$ at $T = 4.2$ K using an *ac* voltage bias $V_0 = 9.0$ mV. The $R_{NL}$ is normalized by $R_{NL}(V_{bg} = 0)$ of 220 kΩ at 5 T (**a**) and 106 kΩ at 3 T (**b**). The $\mathcal{R}_{NL}$ emphasizes the giant nonlocality at larger $|V_{bg}|$ as compared to $R_{NL}$ which drops much faster with $|V_{bg}|$ due to the trivial drop in $\rho_{xx}(|V_{bg}|)$. The $R_{2p}$ shows quantization at QH plateaus as indicated by the dashed lines. **c**, $\mathcal{R}_{NL}$ vs. $V_{bg}$ at $B = 1$, 1.2, and 1.4 T with dashed guide-to-the-eye envelope curves. **d**, $\mathcal{R}_{NL}$ data with envelopes of 1 and 1.2 T data normalized to that of 1.4 T illustrating that the nonlocal mechanism is described by a smooth and continuous envelope function.

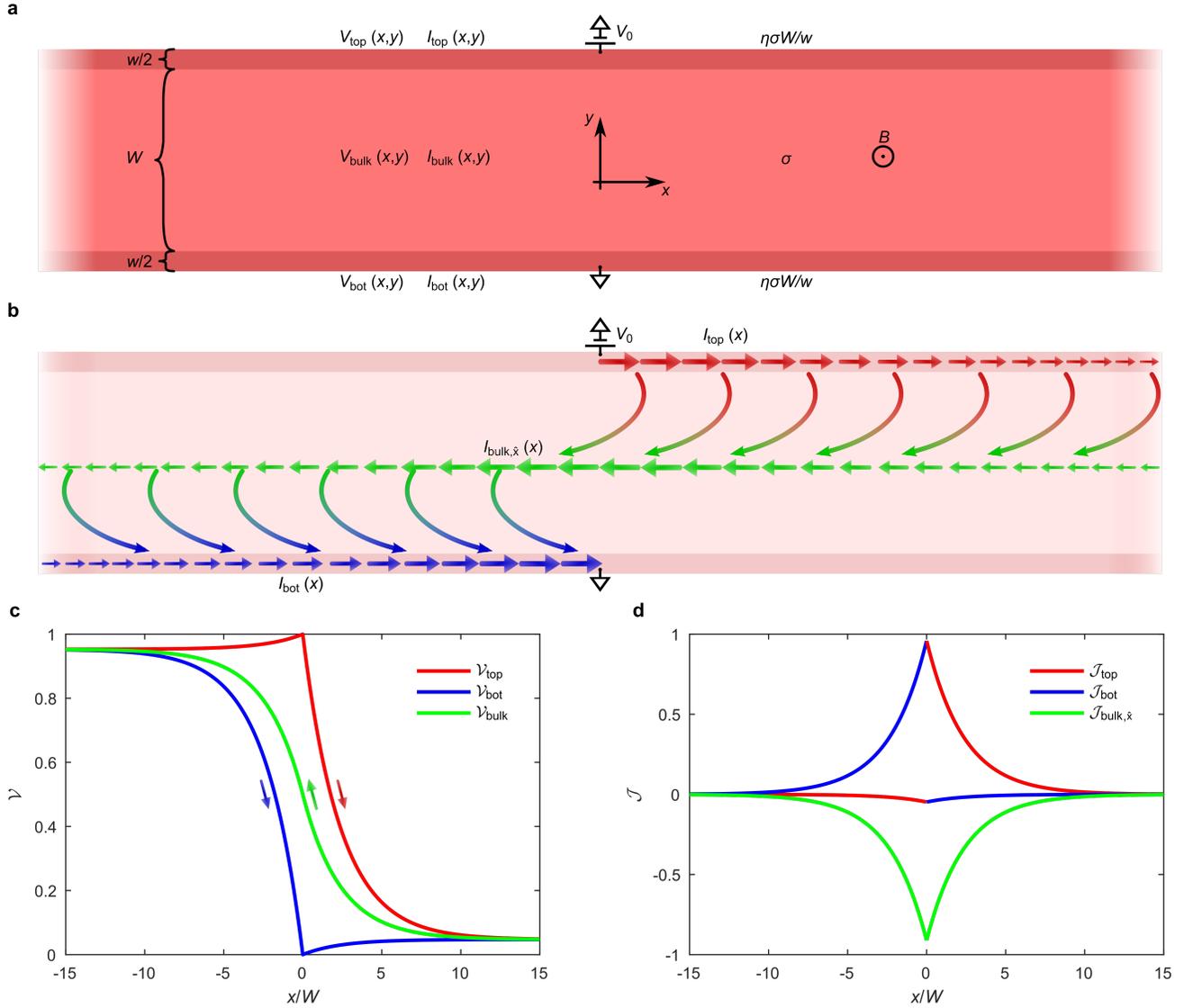

**Extended Data Fig. 2. Analytic solution of nonlocal transport in infinite strip with edge accumulation. a**, Schematic drawing of the sample consisting of a bulk strip of width $W$ described by conductivity tensor $\sigma$ and two edge strips of width $w/2$ with conductivity $\eta\sigma W/w$. All the analytic solutions are derived in the limit $w \to 0$. **b**, Schematic drawing of the current flow at elevated field in presence of edge accumulation. The current flows from the source predominantly along the top-right edge in $\hat{x}$ direction (red) and gradually leaks into the bulk where it reverts its direction and flows in the $-\hat{x}$ direction (green) to the left side of the sample. The current is then gradually absorbed by the bottom edge where it reverts its direction again and flows to the drain (blue). Note that the edge currents (red and blue) flow "downstream" along the potential drop (red and blue arrows in **c**), while the $x$ component of the bulk current, $\mathcal{I}_{bulk,\hat{x}}(x)$, flows against the potential drop (green arrow in **c**). **c**, Calculated normalized potentials $\mathcal{V}_{top}(x)$ (red) and $\mathcal{V}_{bot}(x)$ (blue) along the top and bottom edges and $\mathcal{V}_{bulk}(x, y = 0)$ (green) for the case of $\eta = 0.2$ and $\theta = 26$ ($\lambda = 2.4W$). **d**, The corresponding normalized currents $\mathcal{I}_{top}(x)$ (red), $\mathcal{I}_{bot}(x)$ (blue), and the $x$ component of the bulk current integrated over the strip width, $\mathcal{I}_{bulk,\hat{x}}(x)$, (green) showing the flow pattern described schematically in **b**.

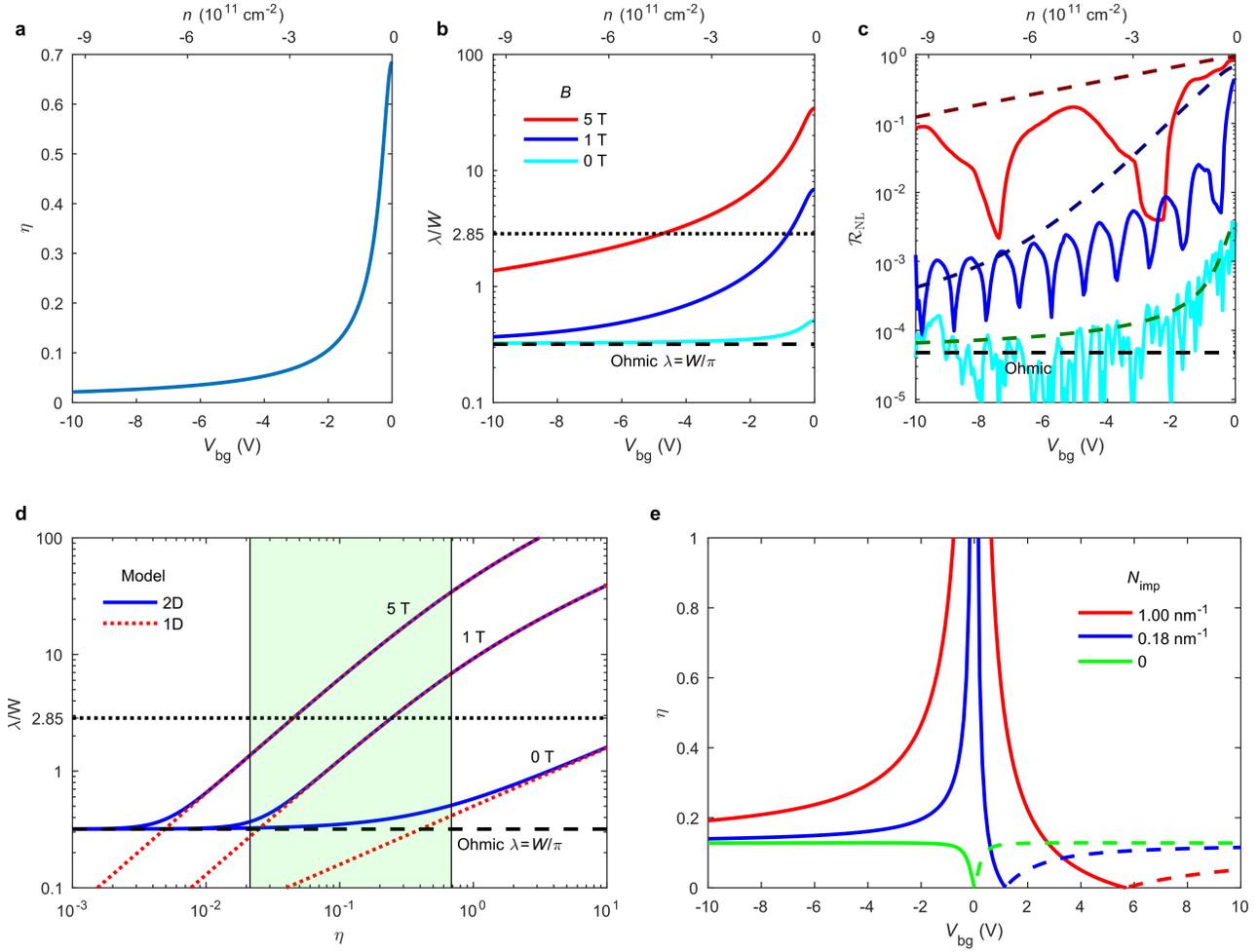

**Extended Data Fig. 3**. **Edge accumulation ratio $\eta$ and nonlocal decay length $\lambda$ behavior. a**, Plot of the calculated edge accumulation ratio $\eta(V_{bg})$ for hole doping with $P_e = 1.8\times10^8$ m$^{-1}$ and $p_{dis} = 2.9\times10^{10}$ cm$^{-2}$ in the limit of $w = 0$. **b**, The corresponding nonlocal decay length $\lambda$ given by Eq. 1 vs. $V_{bg}$ at $B = 0$, 1, and 5 T. The dashed line shows $\lambda/W = 1/\pi$ in the ohmic regime and the dotted line shows the distance to our nonlocal contacts $x/W = 2.85$. **c**, The corresponding calculated $\mathcal{R}_{NL}(V_{bg})$ (dashed) using $\lambda(V_{bg})$ from **b**, along with the experimental $\mathcal{R}_{NL}(V_{bg})$ at $B = 0$, 1, and 5 T from Fig. 1e. **d**, Calculated $\lambda/W$ vs. $\eta$ as in Fig. 1b but on a log-log scale using the 2D analytic expressions (blue) along with the 1D results (dotted red). The colored region shows the span of $\eta(V_{bg})$ in **a**. **e**, COMSOL numerical calculation of $\eta = |2P_e/Wp_b|$ vs. $V_{bg}$ with edge charge accumulation $P_e$ arising due to combined effects of electrostatic gating of the graphene by the backgate potential $V_{bg}$ and of the negatively charged impurities with line density $N_{imp} = 0.18$ nm$^{-1}$ (blue) and 1 nm$^{-1}$ (red) along the edges. A pronounced asymmetry is observed with a faster decay of $\eta$ and of the corresponding nonlocality upon $n$ bulk doping. $\eta$ vanishes at a specific value of positive $V_{bg}$ at which hole edge doping due to $N_{imp}$ is compensated by electron edge doping due to electrostatic gating by $V_{bg}$. At larger $V_{bg}$ both the edges and the bulk become $n$-doped (dashed curves). For the red and blue curves we have used $p_{dis} = 0$, which results in diverging $\eta$ at CNP due to vanishing of $p_b$ at $V_{bg} = 0$. The green curve shows $\eta(V_{bg})$ for the case of pure electrostatic gating ($N_{imp} = 0$) with $p_{dis} = 2.9\times10^{10}$ cm$^{-2}$. In this case the transport is local at CNP with $\eta = 0$ because $P_e$ vanishes at $V_{bg} = 0$ while $p_b = p_{dis}$ is finite.

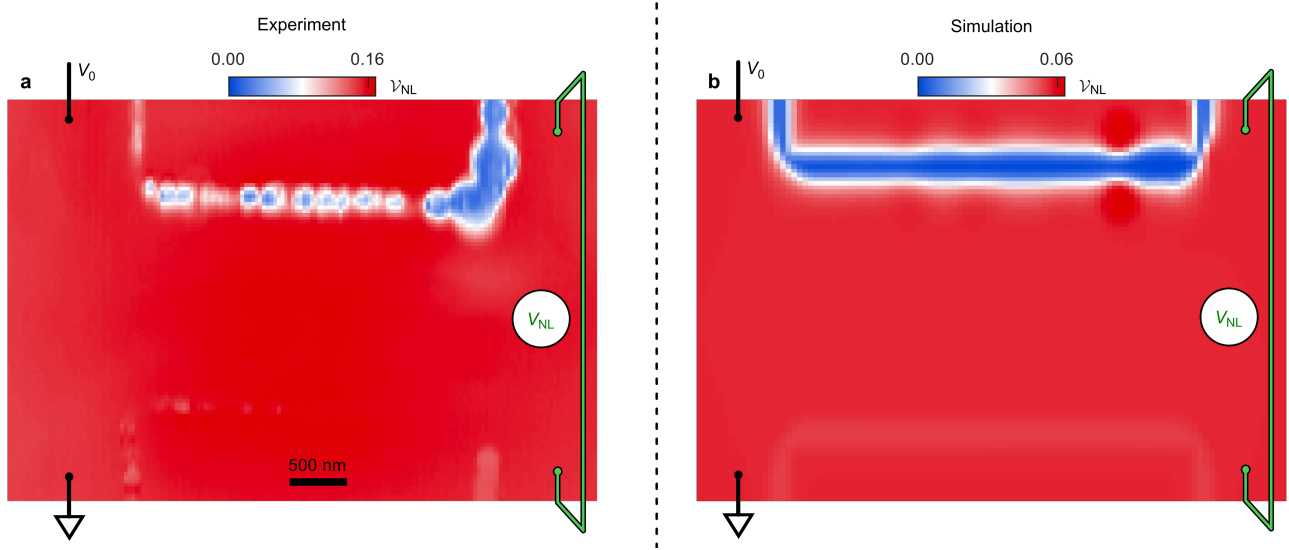

**Extended Data Fig. 4**. **Scanning gate microscopy and simulations away from charge neutrality. a**, Scanning gate image of $\mathcal{V}_{NL} = (V_3 - V_4)/V_0$ at $B = 5$ T, $V_{bg} = -1.6$ V, and $V_{tg} = 8$ V. The graphene device is biased with a voltage $V_0 = 5.5$ mV and the SOT is scanned over the right-hand-side of the sample marked by the white dotted area in Fig. 1a (see Supplementary Video 4 for range of $V_{bg}$ values). At this $V_{bg}$ we estimate the decay length of the edge current to be $\lambda \cong 8W$ (based on Extended Data Fig. 3b and not taking into account the suppression of the edge currents by the edge disorder). Note the large suppression of $\mathcal{V}_{NL}$ by the tip along the top-right edge (blue) and a hardly visible suppression at few locations along the bottom-right edge (light red) in contrast to Fig. 3a that shows a more symmetric suppression close to CNP. **b**, Numerical simulations of $\mathcal{V}_{NL}$ scanning gate signal for $\lambda = 8W$ ($\eta = 0.13$, $V_{bg} = -1.6$ V , $v_{tg} = 2$) with disordered edge accumulation along the top-right edge (with the same disorder function $f(x)$ as in Fig. 3l) and uniform accumulation along the bottom-right edge. A very weak tip-induced suppression of $\mathcal{V}_{NL}$ is visible along the bottom-right edge (light red) consistent with experimental data in (a).

Extended Data Fig. 4 exemplifies the situation in which the decay length $\lambda$ is smaller or comparable to the sample size. In this case the current flows from the source (top) along the top-right edge of the sample and gradually leaks into the bulk where it flows towards the left side of the sample and is collected by the bottom-left edge (outside the scanning widow) towards the drain as shown in Fig. 4f. Hole depletion by the tip potential along the top-right edge cuts the edge current leading to large suppression of $\mathcal{V}_{NL}$ as shown in Figs. 4g,h. In contrast, hole depletion along the bottom-right edge has little effect because most of the edge current has leaked out into the bulk and only a small fraction remains flowing along the bottom-right edge. Scanning the depleting tip across the bulk also has no appreciable effect since the current path is modified only locally without a global effect as shown in Supplementary Movie 8 for an intermediate value of $\lambda = 11.9W$. When $\lambda$ is smaller than the sample size the behavior in finite sample can be approximated by the 2D and 1D analytic solutions presented in the Supplementary Information and in Extended Data Fig. 2.

This case should be contrasted with the situation near CNP where $\lambda$ reaches up to $30W$ at $B = 5$ T (Extended Data Fig. 3b) and becomes substantially larger than the sample size of about $10W$. This case is not captured by the analytical models of infinite sample. For such large $\lambda$ the edge currents flow all the way from source to drain with little decay as discussed in Methods and shown by the numerical simulations in Supplementary Videos 6 and 7. As a result, hole depletion by the tip gives rise to large suppression of $\mathcal{V}_{NL}$ along both the top and bottom edges of the sample as observed experimentally in Fig. 3a and shown by the numerical simulations in Fig. 3j for $\lambda = 30W$. The observed transition from the more symmetric transport along the top and bottom edges near CNP to the asymmetric behavior upon bulk doping is consistent with the model.

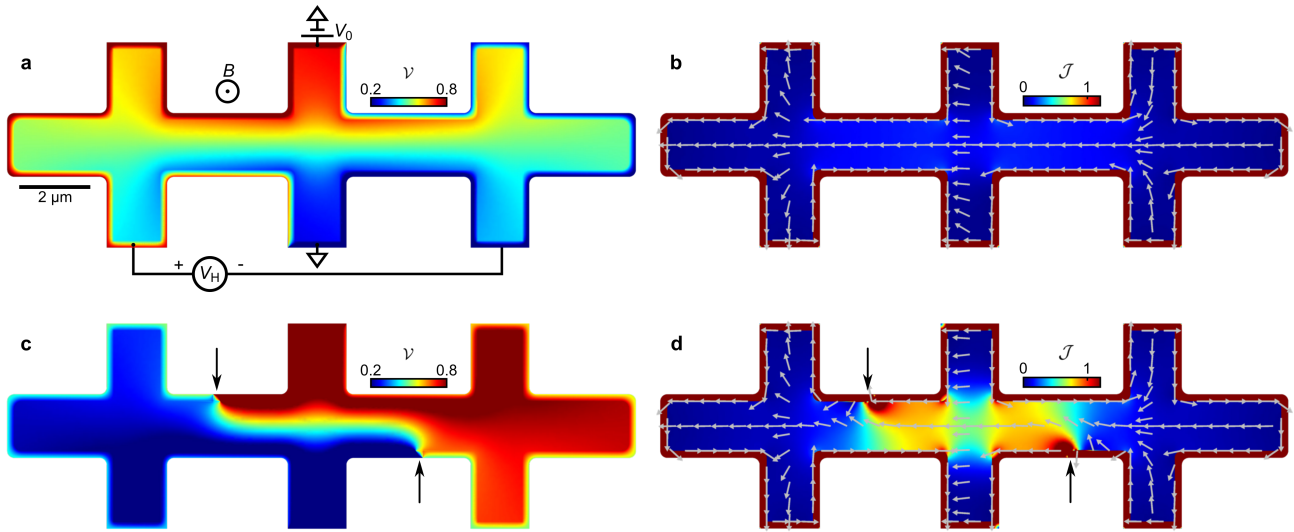

**Extended Data Fig. 5**. **Inversion of the Hall voltage due to edge charge disorder.** Numerical simulations of the normalized potential $\mathcal{V}$ (left column) and of the magnitude of the normalized current density $\mathcal{J}$ (right column) at $B = 5$ T and $\lambda = 30W$ ($\eta = 0.58$, $V_{bg} = -0.2$ V). **a-b**, Uniform edge accumulation giving rise to strong nonlocal transport and positive Hall voltage $V_H$ as defined in **a** (potential at the left contact is higher than at the right contact). **c-d**, The edge charge accumulation is suppressed to 10% of its original value ($f(x) = 0.1$) at two points along the edge marked by the arrows. A significant part of the edge current is diverted into the bulk (**d**) and the potential distribution is altered markedly (**c**) leading to inversion of the Hall voltage (the potential at the left contact is lower than at the right contact).

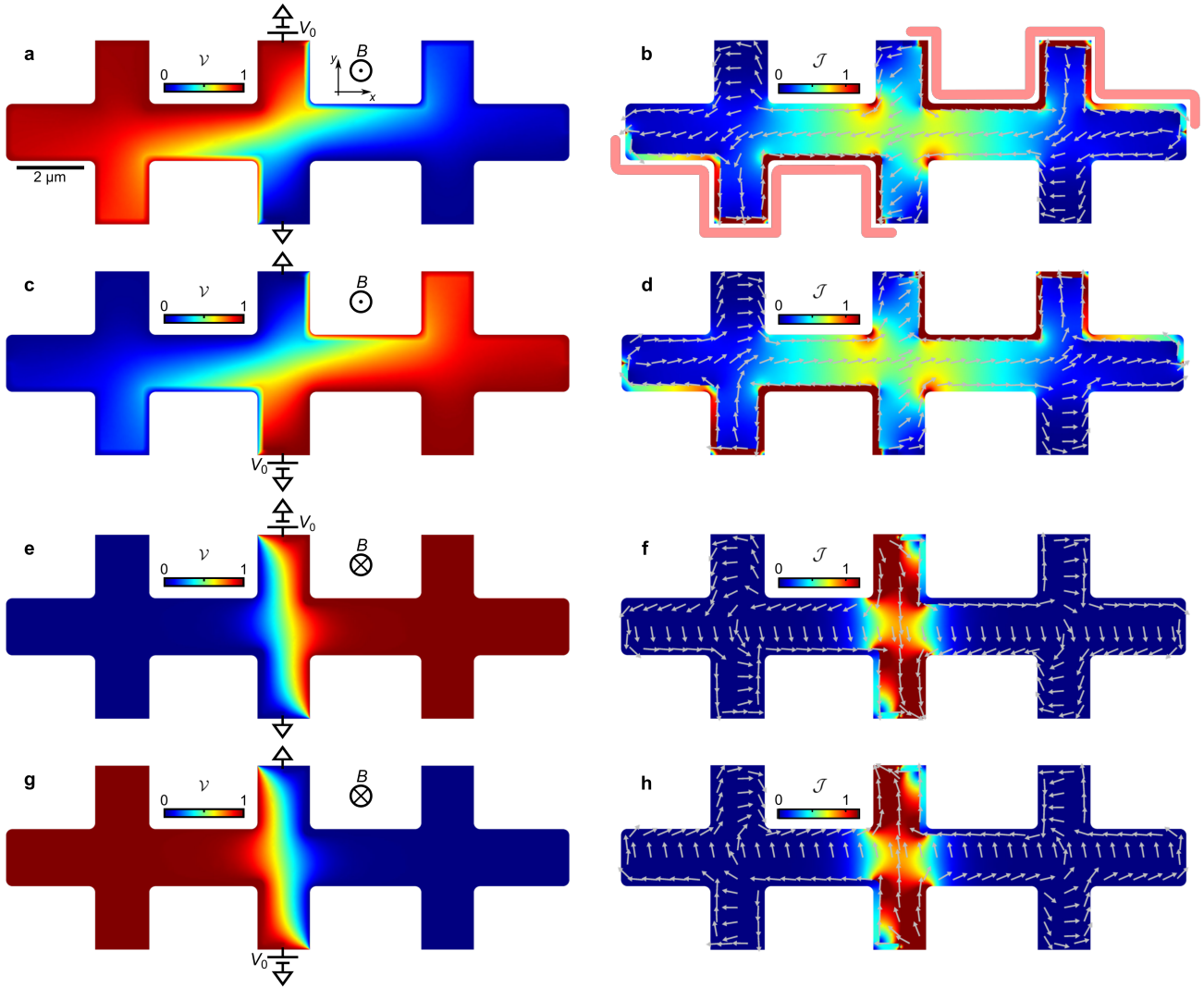

**Extended Data Fig. 6. Field-orientation-dependent nonlocal transport in presence of nonuniform edge accumulation.** Numerical simulations of the normalized potential $\mathcal{V}$ (left column) and of the magnitude of the normalized current density $\mathcal{J}$ (right column) at $B = \pm 4$ T and $\lambda = 2.54W$ ($\eta = 0.05$, $V_{bg} = -4$ V) for the case of charge edge accumulation being present only in the top-right and bottom-left quadrants of the sample indicated by the pink outlines in **b**. **a-b**, $B = 4$ T and $V_0$ applied to the top contact. Highly nonlocal transport is observed similar to the case presented in Figs. 4e,f with charge accumulation along the entire edges. **c-d**, Same as **a-b** but with $V_0$ applied to the bottom contact. The polarities of the potentials and the currents are flipped but the spatial distributions remain the same. **e-f**, $B = -4$ T and $V_0$ applied to the top contact. The transport becomes local resembling the ohmic case in Figs. 4c,d in absence of edge accumulation. **g-h**, Same as **e-f** but with $V_0$ applied to the bottom contact. The transport remains local with flipped current and potential polarities. This nonuniform edge accumulation exemplifies the strong field-orientation dependence of the nonlocal transport that can arise in presence of edge accumulation disorder.

# Supplementary information

**Solution of 2D analytic model**

The electric flow of charge carriers in a strip infinite in the $x$ direction and of width $W$ in the $y$ direction, is determined by the solution of continuity equation $\nabla \cdot \mathbf{J} = 0$, where the current density obeys Ohm's law: $\mathbf{J} = \sigma \mathbf{E} = -\sigma \nabla V$. In a uniform strip the electric potential thus follows Laplace's equation $\nabla^2 V = 0$, resulting in $R_{NL}(x) = \frac{\rho_{xx}}{\pi} \ln\left[\frac{\cosh(\pi x/W)+1}{\cosh(\pi x/W)-1}\right]$, which reduces to $R_{NL}(x) = \frac{4\rho_{xx}}{\pi} e^{-\pi|x|/W}$ for $|x| \gg W$ [1]. However, in the presence of charge accumulation along the boundaries, the conductivity tensor $\sigma$ is no longer uniform, hence the electric field obeys a more complex equation:

$$\sigma_{xx}\nabla^2 V + (\partial_x \sigma_{xx} - \partial_y \sigma_{xy})\partial_x V + (\partial_y \sigma_{xx} + \partial_x \sigma_{xy})\partial_y V = 0. \tag{S1}$$

A general solution of this equation is difficult. Hence, we consider a simplified case of a strip of width $W$ with uniform bulk hole carrier concentration $p_b = -CV_{bg}$ surrounded by two edge strips of width $w/2$ with uniform but higher carrier concentration $p_e = p_b + 2P_e/w$, where $P_e$ is the excess line charge in each edge strip (see Extended Data Fig. 2a). The corresponding conductivity tensor is thus:

$$\sigma(x,y) = \begin{cases} \sigma & |y| \leq \frac{W}{2} \\ \eta \frac{W}{w} \sigma & \frac{W}{2} \leq |y| \leq \frac{W}{2} + \frac{w}{2}, \\ 0 & |y| \geq \frac{W}{2} + \frac{w}{2} \end{cases} \tag{S2}$$

where $\eta = 2\left(P_e + \frac{w}{2} p_b\right)/W p_b$ is the ratio of the total edge to bulk conductance at zero field in local configuration. As a result, Laplace's equation holds separately in each of the regions described by potential profiles $V_{bulk}$, $V_{top}$, and $V_{bot}$, corresponding to the bulk, top edge, and bottom edge strips respectively. The boundary conditions which uniquely define this problem are:

1. Current continuity along the top and bottom edges determined by the current $I_0$ applied to the source and drain point contacts:

$$J_y\left(x, \frac{W+w}{2}\right) = \left[\sigma_{xy}(y)\partial_x V_{top}(x,y) - \sigma_{xx}(y)\partial_y V_{top}(x,y)\right]_{y=\frac{W+w}{2}} = -I_0 \delta(x)$$
$$J_y\left(x, -\frac{W+w}{2}\right) = \left[\sigma_{xy}(y)\partial_x V_{bot}(x,y) - \sigma_{xx}(y)\partial_y V_{bot}(x,y)\right]_{y=-\frac{W+w}{2}} = -I_0 \delta(x) \tag{S3}$$

2. Continuous electric potential throughout the system:

$$V_{top}\left(x, \frac{W}{2}\right) = V_{bulk}\left(x, \frac{W}{2}\right)$$
$$V_{bot}\left(x, -\frac{W}{2}\right) = V_{bulk}\left(x, -\frac{W}{2}\right) \tag{S4}$$

3. Current continuity at the step in the conductivity:

$$\sigma_{xy}\left(\eta \frac{W}{w} - 1\right) \partial_x V_{bulk}|_{y=\frac{W}{2}} = \lim_{\epsilon \to 0} \left[\eta \frac{W}{w} \sigma_{xx} \partial_y V_{top}|_{y=\frac{W}{2}+\epsilon} - \sigma_{xx} \partial_y V_{bulk}|_{y=\frac{W}{2}-\epsilon}\right]$$
$$\sigma_{xy}\left(\eta \frac{W}{w} - 1\right) \partial_x V_{bulk}|_{y=-\frac{W}{2}} = \lim_{\epsilon \to 0} \left[\sigma_{xx} \partial_y V_{bulk}|_{y=-\frac{W}{2}+\epsilon} - \eta \frac{W}{w} \sigma_{xx} \partial_y V_{bot}|_{y=-\frac{W}{2}-\epsilon}\right] \tag{S5}$$

In the simplifying limit $w \to 0$, Eqs. (S1-S5) are solved for the potential profiles in the three regions resulting in the following integrals:

$$V_{top}(x) = 2I_0\rho_{xx}\int_{-\infty}^{\infty}\frac{dk}{2\pi}\frac{e^{ikx}}{k}\left[\frac{\eta W(\theta^2+1)k+2\tanh\left(\frac{kW}{2}\right)+2i\theta}{(\eta W)^2(\theta^2+1)k^2+4W\eta k\coth(kW)+4}\right], \tag{S6a}$$

$$V_{bot}(x) = -2I_0\rho_{xx}\int_{-\infty}^{\infty}\frac{dk}{2\pi}\frac{e^{ikx}}{k}\left[\frac{\eta W(\theta^2+1)k+2\tanh\left(\frac{kW}{2}\right)-2i\theta}{(\eta W)^2(\theta^2+1)k^2+4W\eta k\coth(kW)+4}\right], \tag{S6b}$$

$$V_{bulk}(x,y) = -2I_0\rho_{xx}\int_{-\infty}^{\infty}\frac{dk}{2\pi}\frac{e^{ikx}}{k}\left[\frac{\eta W(\theta^2+1)k\frac{\sinh(ky)}{\sinh(kW/2)}+2\frac{\sinh(ky)}{\cosh(kW/2)}-2i\theta\frac{\cosh(ky)}{\cosh(kW/2)}}{(\eta W)^2(\theta^2+1)k^2+4W\eta k\coth(kW)+4}\right]. \tag{S6c}$$

Analytic solutions of these integrals can be obtained in the following limits.

**Limit of $\eta\sqrt{\frac{\theta^2+1}{\eta+1}} \gg 1$**

In this limit the integrals in Eqs. (S6) can be solved by approximating the integrands around $\left|\frac{kW}{2}\right| \ll 1$, resulting in:

$$\mathcal{V}_{top}(x) = V_{top}(x)/V_0 = \frac{e^{-|x|/\lambda}}{2}\left[1+\frac{\eta\theta}{\eta+1}\frac{\sqrt{(\eta+1)(\theta^2+1)}}{\eta(\theta^2+1)+1}\text{sgn}(x)\right] - \frac{1}{2}\frac{\eta\theta}{\eta+1}\frac{\sqrt{(\eta+1)(\theta^2+1)}}{\eta(\theta^2+1)+1}\text{sgn}(x), \tag{S7a}$$

$$\mathcal{V}_{bot}(x) = -\frac{e^{-|x|/\lambda}}{2}\left[1-\frac{\eta\theta}{\eta+1}\frac{\sqrt{(\eta+1)(\theta^2+1)}}{\eta(\theta^2+1)+1}\text{sgn}(x)\right] - \frac{1}{2}\frac{\eta\theta}{\eta+1}\frac{\sqrt{(\eta+1)(\theta^2+1)}}{\eta(\theta^2+1)+1}\text{sgn}(x), \tag{S7b}$$

$$\mathcal{V}_{bulk}(x,y) = \frac{e^{-|x|/\lambda}}{2}\left[\frac{2y}{W}+\frac{\eta\theta}{\eta+1}\frac{\sqrt{(\eta+1)(\theta^2+1)}}{\eta(\theta^2+1)+1}\text{sgn}(x)\right] - \frac{1}{2}\frac{\eta\theta}{\eta+1}\frac{\sqrt{(\eta+1)(\theta^2+1)}}{\eta(\theta^2+1)+1}\text{sgn}(x), \tag{S7c}$$

$$\lambda = \frac{\eta W}{2}\sqrt{\frac{\theta^2+1}{\eta+1}}, \tag{S7d}$$

$$R_{2p} = \rho_{xx}\frac{\eta(\theta^2+1)+1}{\eta\sqrt{(\eta+1)(\theta^2+1)}}. \tag{S7e}$$

The current density profiles along the edge regions and in the bulk can be calculated next from Ohm's law with the proper conductivity in each region $\boldsymbol{j}_i = -\sigma^i\nabla V = -\frac{\eta W/w}{(\theta^2+1)\rho_{xx}}\begin{pmatrix}1 & -\theta \\ \theta & 1\end{pmatrix}\nabla V$. Since the $w \to 0$ limit and the derivative with respect to $y$ do not commute, the latter should be performed first. Furthermore, upon taking the limit $w \to 0$ the current density in the edge regions diverges, whereas the total current $\frac{w}{2}(\boldsymbol{j}\cdot\hat{x})$ remains finite. The edge currents and the current density in the bulk are thus readily obtained in their integral form:

$$\mathcal{I}_{top}(x) = \frac{\boldsymbol{j}_{top}(x)\cdot\hat{x}}{I_0}\frac{w}{2} = -i\eta W\int_{-\infty}^{\infty}\frac{dk}{2\pi}e^{ikx}\left[\frac{\eta W(\theta^2+1)k+2\tanh\left(\frac{kW}{2}\right)+2i\theta}{(\eta W)^2(\theta^2+1)k^2+4W\eta k\coth(kW)+4}\right], \tag{S8a}$$

$$\mathcal{I}_{bot}(x) = \frac{\boldsymbol{j}_{bot}(x)\cdot\hat{x}}{I_0}\frac{w}{2} = i\eta W\int_{-\infty}^{\infty}\frac{dk}{2\pi}e^{ikx}\left[\frac{\eta W(\theta^2+1)k+2\tanh\left(\frac{kW}{2}\right)-2i\theta}{(\eta W)^2(\theta^2+1)k^2+4W\eta k\coth(kW)+4}\right], \tag{S8b}$$

$$\mathcal{J}_{bulk}(x,y) = \frac{\boldsymbol{j}_{bulk}(x)}{I_0}W = -2W\int_{-\infty}^{\infty}\frac{dk}{2\pi}e^{ikx}\begin{pmatrix}\frac{i\eta Wk\frac{\sinh(ky)}{\sinh(kW/2)}+\theta\eta Wk\frac{\cosh(ky)}{\sinh(kW/2)}+2i\frac{\sinh(ky)}{\cosh(kW/2)}}{(\eta W)^2(\theta^2+1)k^2+4W\eta k\coth(kW)+4} \\ \frac{-i\eta Wk\frac{\sinh(ky)}{\sinh(kW/2)}+\eta Wk\frac{\cosh(ky)}{\sinh(kW/2)}+2\frac{\cosh(ky)}{\cosh(kW/2)}}{(\eta W)^2(\theta^2+1)k^2+4W\eta k\coth(kW)+4}\end{pmatrix}. \tag{S8c}$$

These Fourier transforms are once again solved by approximating the integrands for $\left|\frac{kW}{2}\right| \ll 1$ to give

$$\mathcal{I}_{top}(x) = \frac{e^{-|x|/\lambda}}{2} \left[ \frac{\eta(\theta^2+1)+1}{\eta(\theta^2+1)} \text{sgn}(x) + \frac{\theta}{\sqrt{(\eta+1)(\theta^2+1)}} \right], \tag{S9a}$$

$$\mathcal{I}_{bot}(x) = -\frac{e^{-|x|/\lambda}}{2} \left[ \frac{\eta(\theta^2+1)+1}{\eta(\theta^2+1)} \text{sgn}(x) - \frac{\theta}{\sqrt{(\eta+1)(\theta^2+1)}} \right], \tag{S9b}$$

$$\boldsymbol{J}_{bulk}(x,y) = -\frac{e^{-|x|/\lambda}}{\eta(\theta^2+1)} \begin{pmatrix} -\frac{2y}{W}\frac{(\eta+1)}{\eta} \text{sgn}(x) + \theta\eta\sqrt{\frac{\theta^2+1}{\eta+1}} \\ \frac{2y}{W}\theta\,\text{sgn}(x) + \sqrt{(\eta+1)(\theta^2+1)} \end{pmatrix}. \tag{S9c}$$

Note that the 1D currents in the edge regions obey continuity relations with respect to the bulk 2D current density, i.e. $\partial \mathcal{I}_{top}/\partial x = \boldsymbol{J}_{bulk,\hat{y}}(x, W/2)$ and $\partial \mathcal{I}_{bot}/\partial x = -\boldsymbol{J}_{bulk,\hat{y}}(x, -W/2)$, where $\boldsymbol{J}_{bulk,\hat{y}}$ is the component of the bulk current density flowing in the $\hat{y}$ direction. Note that in contrast to the zero field case, the bulk current flows predominantly in $\hat{x}$ direction, $\boldsymbol{J}_{bulk,\hat{x}} \gg \boldsymbol{J}_{bulk,\hat{y}}$, and is essentially $y$ independent. This rotation of the bulk current direction with increasing $\theta$ or $\eta$ is clearly seen in Supplementary Videos 6 and 7.

The above results apply also to the case of $\theta = 0$ (for $\eta \gg 1$) giving rise to $\lambda = \frac{\eta W}{2\sqrt{\eta+1}}$. In the Methods we summarize these results valid for $\eta\sqrt{\frac{\theta^2+1}{\eta+1}} \gg 1$ and with the voltage set to zero at the drain contact, $V_{bot}(0, -W/2) = 0$. The corresponding current and voltage profiles are presented in Extended Data Figs. 2c,d.

**Limit of $\eta\sqrt{\frac{\theta^2+1}{\eta+1}} \not\gg 1$**

The general case $\eta\sqrt{\frac{\theta^2+1}{\eta+1}} \not\gg 1$ results in two qualitatively dissimilar solutions depending whether $x \ll \lambda$ or $x \gg \lambda$. We first discuss the far limit ($x \gg \lambda$), for which the integrals in Eq. (S6) are amenable to a contour integration in the upper half-plane as they are completely dominated by the pole closest to the real axis. The location of this pole is given by the first complex valued $iz$ which solves

$$-(\eta W)^2(\theta^2+1)z^2 + 4W\eta \cot(Wz)z + 4 = 0. \tag{S10}$$

Here, $iz$ is purely imaginary, with the limits $z \to \frac{\pi}{W}(1-\eta)$ for $\theta, \eta \ll 1$ and $z \to \frac{2}{\eta W}\sqrt{\frac{\eta+1}{\theta^2+1}}$ for $\theta \gg 1$. We emphasize that it is necessary to match the term of first order in $\eta$ to obtain reliable results. An approximate solution for the transcendental equation is given by $z = 1/\lambda$ with

$$\lambda = \frac{W}{\pi}\sqrt{1 + 2\eta\left[1 + \frac{\pi^2\eta}{4}\left(\frac{1+\eta(\theta^2+1)}{\eta+1}\right)^2\right] \Big/ \left[1 + 2\eta\frac{1+\eta(\theta^2+1)}{\eta+1}\right]}, \tag{S11}$$

as presented in Eq. (1) in the main text, which recovers the asymptotics of the true solution for both large and small values of $\lambda$. Taking the residue in Eq. (S6), one then obtains

$$V_{top}(x) - V_{bot}(x) = 2I_0\rho_{xx}\frac{e^{-x/\lambda}}{W/\lambda}\left[\frac{\eta(\theta^2+1)W/\lambda + 2\tan(W/2\lambda)}{\eta(\theta^2+1)W/\lambda - 2\cot(W/\lambda) + 2W/\lambda\sin^2(W/\lambda)}\right]. \tag{S12}$$

In the limit $\eta\sqrt{\frac{\theta^2+1}{\eta+1}} \ll 1$ one thus obtains $\lambda = \frac{W}{\pi}(1+\eta)$ and $R_{NL}(x) = \frac{4}{\pi}\rho_{xx}e^{-|x|/\lambda}$ in the far limit $x \gg \lambda$.

Since we analyze the experimental $R_{NL}$ by normalizing by $R_{2p}$, we also have to derive $R_{NL}$ in the $x \ll \lambda$ limit to obtain $R_{2p} = R_{NL}(x \to 0)$. In the limit $\eta\sqrt{\frac{\theta^2+1}{\eta+1}} \gg 1$, $R_{NL}(x) = R_{2p}e^{-|x|/\lambda}$ at all $x$ with $R_{2p} = \rho_{xx}\sqrt{\frac{\theta^2+1}{\eta+1}}$. In the ohmic regime with $\eta = 0$, in contrast, $R_{NL}(x) = \frac{\rho_{xx}}{\pi}\ln\left[\frac{\cosh(\pi x/W)+1}{\cosh(\pi x/W)-1}\right]$, which has a logarithmic divergence at $x \to 0$ due to zero size contacts. Interestingly, this divergence is eliminated for any finite $\eta$ or $\theta$. To derive $R_{2p}$ in this limit one needs to resolve faithfully also the large values of $k$ in Eq. (S6). To this end, we substitute $k \to \tilde{k}/\eta$ and approximate for small $\eta$, arriving at

$$R_{2p} = (V_{top}(0) - V_{bot}(0))/I_0 = 2\rho_{xx}\int_0^\infty \frac{d\tilde{k}}{\pi}\frac{1}{W\tilde{k}+2\eta}\left[\frac{(\theta^2+1)W\tilde{k}+2}{(\theta^2+1)(W\tilde{k})^2+4W\tilde{k}+4}\right] =$$
$$2\rho_{xx}\frac{2\theta\arcsec(\sqrt{\theta^2+1})-\ln((\theta^2+1)\eta^2)}{2\pi},$$
(S13)

where a cutoff was inserted at the infrared scale $2\eta$. For finite magnetic fields, it is sufficient to use $R_{2p}$ as given by Eq. (S13) as the normalization for all values $\eta$ of interest. On the other hand, at $\theta = 0$ the logarithmic enhancement of $R_{2p}$ is substantially smaller, necessitating an interpolation that smoothly connects $R_{2p}$ for small $\eta$ in Eq. (S13) with the value of $R_{2p}$ at large $\eta$ [Eq. (S7e)]. To this end, we generalize the logarithmic piece so that it decays for $\eta \to \infty$ and adjust one free parameter $c$, to find

$$R_{2p} = 2\rho_{xx}\left(\frac{1}{2\pi}\ln\frac{\cosh(\eta)+1}{\cosh(\eta)-1} + \frac{1}{2\sqrt{\eta}+c/\eta}\right), \quad c \approx 4.$$
(S14)

We point out that there is an additional intermediate regime for $x \ll \lambda$ and $\eta\sqrt{\frac{\theta^2+1}{\eta+1}} \ll 1$ where the previously discussed approximations for $R_{NL}(x)$ fail, but it falls outside the parameter regime covered in the present work.

Since $R_{2p}$ is not well defined for $\eta = 0$, we have used $\eta = 0.01$ for plotting the dashed curves of ohmic $\mathcal{R}_{NL} = R_{NL}/R_{2p}$ in Fig. 1e and Extended Data Fig. 3c, which corresponds to the extrapolated $\eta$ value at $V_{bg} = -20$ V with corresponding $R_{2p} \cong 5.3\rho_{xx}$.

Note that our samples are in the ballistic regime with the mean free path comparable to the sample size. Therefore, the concept of ohmic conductivity cannot be readily applied at low fields and more elaborate models are needed. At elevated fields, however, the magnetic length becomes the relevant characteristic length scale of the problem. At 5 T the magnetic length is about 10 nm, about three orders of magnitude smaller than the sample size, justifying the use of the ohmic conductivity description.

**Solution of 1D analytic model**

Given the current behavior in the 2D model, we can construct a simpler model of nonlocal transport at elevated fields. The top and bottom edges of the sample are modeled by two infinite 1D wires with resistance per unit length of $R_{edge} = 2\frac{\rho_{xx}}{\eta W}$ (based on definition of bulk to edge conductance ratio $\eta = \frac{\rho_{xx}/W}{R_{edge}/2}$) carrying currents $I_{top}(x)$ and $I_{bot}(x)$ given by Ohm's relation

$$-\frac{dV_{top}(x)}{dx} = 2\frac{\rho_{xx}}{\eta W}I_{top}(x)$$
$$-\frac{dV_{bot}(x)}{dx} = 2\frac{\rho_{xx}}{\eta W}I_{bot}(x)$$
(S15)

The leakage of the current between the edges and the bulk is described by the continuity equations

$$\frac{dI_{top}(x)}{dx} = j_{bulk,\hat{y}}\left(x, \frac{W}{2}\right)$$
$$\frac{dI_{bot}(x)}{dx} = -j_{bulk,\hat{y}}\left(x, -\frac{W}{2}\right), \tag{S16}$$

where $j_{bulk,\hat{y}}\left(x, \pm\frac{W}{2}\right)$ are given by Ohm's law in the bulk

$$j_{bulk,\hat{y}}\left(x, \frac{W}{2}\right) = \sigma_{xy}\frac{dV_{top}(x)}{dx} - \sigma_{xx}\frac{V_{top}(x)-V_{bot}(x)}{W}$$
$$j_{bulk,\hat{y}}\left(x, -\frac{W}{2}\right) = \sigma_{xy}\frac{dV_{bot}(x)}{dx} - \sigma_{xx}\frac{V_{top}(x)-V_{bot}(x)}{W}, \tag{S17}$$

Here we have assumed that the potential difference between $V_{top}(x)$ to $V_{bot}(x)$ drops linearly across the bulk in the $y$ direction, in accordance with Eq. (S7c) in the 2D model.

By substituting Eqs. (S16) and (S17) into Eq. (S15) a set of coupled second order differential equations is obtained for $V_{top}$ and $V_{bot}$

$$-\frac{\eta W(\theta^2+1)}{2}\frac{d^2V_{top}(x)}{dx^2} = \theta\frac{dV_{top}(x)}{dx} - \frac{V_{top}(x)-V_{bot}(x)}{W}, \tag{S18a}$$

$$\frac{\eta W(\theta^2+1)}{2}\frac{d^2V_{bot}(x)}{dx^2} = \theta\frac{dV_{bot}(x)}{dx} - \frac{V_{top}(x)-V_{bot}(x)}{W}, \tag{S18b}$$

with boundary conditions

$$V_{top}(0) = V_0 \;;\; V_{bot}(0) = 0 \;;\; (V_{top} - V_{bot}) \xrightarrow{|x|\to\infty} 0. \tag{S19}$$

These equations are readily solved using ansatz solutions

$$V_{top}(x) = Ae^{-x/\lambda} + B, \tag{S20a}$$

$$V_{bot}(x) = Ce^{-x/\lambda} + D, \tag{S20b}$$

where $A, B, C, D$ and $\lambda$ are coefficients that need to be determined for positive and negative $x$. Substituting Eqs. (S20a,b) into Eqs. (S18a,b) with Eq. (S19) boundary conditions, we find the potential profiles for $x \geq 0$:

$$\mathcal{V}_{top}(x \geq 0) = \frac{1}{2}\left(1 + \frac{\theta}{\sqrt{(\eta+1)(\theta^2+1)-1}}\right)\left(e^{-x/\lambda} - 1\right) + 1, \tag{S21a}$$

$$\mathcal{V}_{bot}(x \geq 0) = \frac{1}{2}\left(1 - \frac{\theta}{\sqrt{(\eta+1)(\theta^2+1)-1}}\right)\left(1 - e^{-x/\lambda}\right), \tag{S21b}$$

and for $x < 0$:

$$\mathcal{V}_{top}(x < 0) = \frac{1}{2}\left(1 - \frac{\theta}{\sqrt{(\eta+1)(\theta^2+1)-1}}\right)\left(e^{x/\lambda} - 1\right) + 1, \tag{S22a}$$

$$\mathcal{V}_{bot}(x < 0) = \frac{1}{2}\left(1 + \frac{\theta}{\sqrt{(\eta+1)(\theta^2+1)-1}}\right)\left(1 - e^{x/\lambda}\right), \tag{S22b}$$

with

$$\lambda = \frac{\eta W}{2}\frac{\theta^2+1}{\sqrt{(\eta+1)(\theta^2+1)-1}}, \tag{S23a}$$

$$R_{2p} = \frac{2\rho_{xx}\lambda}{\eta W} = \frac{\rho_{xx}(\theta^2+1)}{\sqrt{(\eta+1)(\theta^2+1)-1}}, \tag{S23b}$$

and corresponding $\mathcal{I}_{top}$ and $\mathcal{I}_{bot}$ given by Eq. (S15) and $\mathcal{I}_{bulk} = -(\mathcal{I}_{top} + \mathcal{I}_{bot})$.

Extended Data Fig. 3d shows the comparison of $\lambda(\eta)$ in the 1D and 2D models for several field values. As long as $\eta\sqrt{\frac{\theta^2+1}{\eta+1}} \gg 1$, the 1D model coincides with the 2D results, but it fails when $\lambda$ approaches the ohmic value.

Note, however, that both models deal with infinite strips with point contacts and thus can only provide a general insight into the more realistic situation of finite size samples with finite-width contacts, which we address by COMSOL simulations.

**Magnetic-field-induced unidirectional bulk-edge decoupling**

The 1D model provides a key insight into the origin of the giant nonlocality in terms of field-induced bulk-edge decoupling. In a nonlocal configuration, the coupling strength between the edges and the bulk can be quantified by the leakage rate of the edge current to the bulk. At zero magnetic field Eqs. (S16-17) can be approximated as

$$\frac{\mathrm{d}I_{top}(x)}{\mathrm{d}x} \cong \sigma_{xx} \frac{V_{NL}(x)}{W}. \tag{S24}$$

At zero field the bulk conductivity $\sigma_{xx} = \sigma_0$ is high and hence the conducting edges are essentially shorted to the bulk.

At elevated fields the situation changes as follows. According to Eq. (S15), the edge behaves as a 1D wire with current determined by its magnetic field independent resistivity $R_{edge} = \frac{2\rho_{xx}}{\eta W} = \frac{2}{\eta W \sigma_0}$ and by the electric field along the edge,

$$I_{top}(x) = -\frac{\mathrm{d}V_{top}(x)}{\mathrm{d}x}/R_{edge} = -\frac{\eta W}{2\rho_{xx}} \frac{\mathrm{d}V_{top}(x)}{\mathrm{d}x}. \tag{S25}$$

The leakage into the bulk, in contrast, is strongly magnetic field dependent as described by Eqs. (S16-17). In the limit of $\eta \ll 1$, for $x \geq 0$ it is mainly governed by the Hall current induced by the electric field along the edge and by $\sigma_{xy}$,

$$\frac{\mathrm{d}I_{top}(x)}{\mathrm{d}x} \cong \sigma_{xy} \frac{\mathrm{d}V_{top}(x)}{\mathrm{d}x}, \tag{S26}$$

with $\sigma_{xy} \cong \sigma_0/\theta$. The resulting bulk-edge coupling,

$$\frac{\mathrm{d}I_{top}(x)/\mathrm{d}x}{I_{top}(x)} \cong -\frac{2\sigma_{xy}\rho_{xx}}{\eta W} = -\frac{2}{\eta W \theta}, \tag{S27}$$

is thus proportional to $\sigma_{xy}\rho_{xx} = 1/\theta = 1/\mu B$, giving rise to a strong field-induced bulk-edge decoupling, which is of the key effect that enables the giant nonlocality. The corresponding edge current decay length, $\lambda \cong \frac{\eta W \theta}{2} = \frac{\eta W \mu B}{2}$, grows linearly with magnetic field, which leads to an exponential growth of $R_{NL}$ in an infinite sample. In finite samples, this growth is saturated when $\lambda$ becomes comparable to the sample size.

In zero magnetic field, the strong bulk-edge coupling allows the edge current to partially circumvent the disorder-induced barriers by bypassing them through the bulk. The magnetic-field-induced decoupling, in contrast, is driven by the Hall current and therefore acquires a unidirectional character. Equation (S27) shows that for $x \geq 0$ the top edge leaks the current unidirectionally into the bulk, while for $\eta \ll 1$ the bottom edge carries almost no current. As a result, along the top-right edge, the edge current that leaks into the bulk due to disorder-induced local barriers cannot return to the edge. For $x < 0$ a reverse behavior takes place: the top edge carries almost no current while the bottom edge unidirectionally gutters the current from the bulk into the edge channel. Hence, along the bottom-left edge the current that drains from the bulk into the edge cannot escape back to the bulk. Edge currents that can resonantly tunnel through a disorder-induced barrier, however, will maintain their flow along the edge. This is the reason for the edge currents becoming so sensitive to disorder at elevated fields, a behavior that results in sharp resonant features observed in our scanning gate images.

**COMSOL simulations of nonlocal transport in a finite sample**

The 2D finite element simulations were performed using COMSOL Multiphysics 5.4 that takes into account both edge accumulation and the effect of the tip potential. We used the Electric Currents (EC) module, which receives the conductivity tensor $\sigma(x,y)$ as an input parameter and solves Ohm's law and the continuity equation for the potential $V$ in a defined geometry. The overall dimensions of the simulated device were chosen to be equivalent to the experimental device, with bulk regions of width $W = 1.4$ μm and an edge strip of width $w/2 = 200$ nm surrounding the bulk (the overall width being 1.8 μm as in the experiment). The value of $w$ was found not to affect the results significantly as long as the total excess line charge $P_e$ in the edge strip is kept constant. We have used rather large $w/2 = 200$ nm in order to allow a clear resolution of the edge and bulk regions in the presented figures and videos. Since the current density varies by orders of magnitude across the sample, we render the modulus of the normalized current density $\mathcal{J} = |\boldsymbol{J}|W/I_0$ by the color code with regions of high density $\mathcal{J} > 1.2$ saturated to dark red for clarity. The grey arrows represent the local direction of the current along the edge-bulk interface and along the central lines in the bulk.

Since both edge accumulation and the tip affect the charge carrier density, we modeled the conductivity tensor as $\sigma_{xx}^i = (p_i e\mu)/(1+\theta^2)$ and $\sigma_{yx}^i = \theta\sigma_{xx}^i$, where $p_i$ is the hole carrier density in the bulk region ($i = b$) and in the edge strip ($i = e$). In the bulk $p_b = \sqrt{p_{dis}^2 + (CV_{bg})^2}$, where $p_{dis} = 2.9 \times 10^{10}$ cm$^{-2}$ is a constant charge density which reflects the average bulk charge disorder, which is introduced here to avoid singularities of $\sigma$ at CNP. In the edge strip, the edge accumulation charge density $2P_e/w$ is added to the bulk hole density, thus $p_e = \sqrt{p_{dis}^2 + \left(\frac{2P_e}{w} - CV_{bg}\right)^2}$. The positive tip potential $V_{tg}$ causes local depletion of the holes, which is modeled by taking $p_{bulk}(\boldsymbol{r}) = p_b - p_{tip}(\boldsymbol{r}-\boldsymbol{r_0})$ and $p_{edge}(\boldsymbol{r}) = p_e - p_{tip}(\boldsymbol{r}-\boldsymbol{r_0})$, where $p_{tip}(\boldsymbol{r}-\boldsymbol{r_0}) = v_{tg}p_b e^{-|\boldsymbol{r}-\boldsymbol{r_0}|^2/2G}$, $v_{tg}$ is a unitless parameter controlling the tip's depletion strength and $G = 100$ nm. In order to avoid the formation of $pn$ junctions, which require more involved calculations, only $p$-doping is considered and any negative local carrier density is forced to zero.

A constant $V_0 = 100$ μV was applied to the entire horizontal edge of the top central contact and a zero potential applied to the horizontal edge of the bottom central contact. The rest of the edges obey the boundary condition for the current density, $\hat{n} \cdot \boldsymbol{J} = 0$, where $\hat{n}$ is the unit vector normal to the edges. Experimental values were taken for all remaining parameters.

**Variability between samples, contact configuration, and field orientation**

The nonlocal transport in graphene has been reported to show large variability between different samples, contact configurations, and magnetic field orientation [1,24,29]. Local suppression of the edge accumulation due to disorder diverts current from the edge into the bulk and reduces the amount of current flowing in the rest of the edge (see Supplementary Videos 8 and 9). As a result, the amount of edge current that reaches a nonlocal contact is determined by the weakest edge accumulation point between the source and the contact as revealed by the scanning gate data and simulations in Fig. 3. Consequently, $\mathcal{R}_{NL}$ in different samples or across different contacts is very sensitive to the particular realization of the disorder in the edge charge accumulation. This effect is clearly visible in Supplementary Video 8. When the tip is in the bulk of the sample, the edge accumulation is uniform resulting in a large $\mathcal{V}_{NL}$ that is equal at the right, $\mathcal{V}_3 - \mathcal{V}_4$, and the left, $\mathcal{V}_5 - \mathcal{V}_6$, nonlocal contacts (see layout in Fig. 1a). However, when the tip (or disorder) induces a local suppression in the edge accumulation along the top-right edge, $\mathcal{V}_3 - \mathcal{V}_4$ drops sharply (uniform dark blue potential $\mathcal{V}$ in the right arm) while the nonlocal voltage in the left arm $\mathcal{V}_5 - \mathcal{V}_6$ is hardly affected. An opposite situation occurs when the tip is located along the bottom-left edge: the nonlocal transport and $\mathcal{V}_5 - \mathcal{V}_6$ are strongly suppressed in the left arm (uniform dark red $\mathcal{V}$ in the left arm) while the nonlocal $\mathcal{V}_3 - \mathcal{V}_4$ remains large, explaining the reported large sample-to-sample and contact-to-contact variability.

Disorder in the edge accumulation may create a similarly large difference in the nonlocal resistance between the up and down field orientations. To demonstrate an asymmetric nonlocal magnetic field response, Extended Data Fig. 6 shows an example in which edge accumulation is present only in the top-right and bottom-left quadrants of the sample as indicated by the pink contours in panel 6b. For positive $V_0$ applied to the top contact and positive $B$ (in $\hat{z}$ direction), the transport is highly nonlocal with pronounced $\mathcal{R}_{NL}$ as shown in Extended Data Figs. 6a,b. The same nonlocal behavior with reversed polarity is obtained for the opposite direction of the applied current as shown in panels 6c,d, resulting in the same $\mathcal{R}_{NL}$. Remarkably, when the field orientation is flipped to $-\hat{z}$, the transport becomes close to ohmic with vanishing $\mathcal{R}_{NL}$ for both current orientations as shown in panels 6e-h, similar to the ohmic case with no edge accumulation in Figs. 4c,d (albeit with opposite field orientation). This extreme example clarifies the origin of the magnetic field asymmetry of the nonlocal transport in presence of edge accumulation disorder as reported in [1].

Note that despite the fact that the potential and current distributions are vastly different for the two field orientations in Extended Data Fig. 6, the Onsager relations are preserved [1] and $R_{2p}$ is the same for positive and negative fields. Naively, the longer current paths in the nonlocal case at positive $B$ should have resulted in larger dissipation and hence larger $R_{2p}$ than in the ohmic case at negative $B$. In the ohmic regime, however, most of the dissipation occurs at the hotspots at the current contacts. In the presence of edge accumulation with enhanced edge conductance, the nontopological edge currents spread this dissipation over a characteristic length $\lambda$, thus reducing the power density but preserving the overall dissipation.

**Asymmetry between $p$ and $n$ doping**

In our discussion above, we have assumed that the excess line charge $P_e$ accumulated along the edges is determined by negatively charged impurities with line density $N_{imp}$ and the induced $P_e$ is $V_{bg}$ independent, which is valid near CNP. At larger $|V_{bg}|$ values, however, edge accumulation due to electrostatic gating by the backgate [35,52,68–70] may become important. In order to evaluate the resulting edge accumulation we have performed COMSOL simulations of the charge density distribution across a graphene strip separated by hBN from the backgate at potential $V_{bg}$ with dimensions equivalent to the experimental setup. A 1D line of negatively charged impurities with density of $N_{imp}$ is placed in the plane of graphene at a distance of 1 nm outside the graphene edges. The charge density in graphene induced by $N_{imp}$ and $V_{bg}$ is calculated self-consistently taking into account the Dirac density of states of graphene. The line charge $P_e$ along the edges in excess of the bulk carrier density $p_b$ is then computed vs. $V_{bg}$. The resulting edge accumulation ratio $\eta = |2P_e/Wp_b|$ is plotted in Extended Data Fig. 3e for $N_{imp} = 0.18$ nm$^{-1}$ (blue) and 1 nm$^{-1}$ (red). In the limit of a large negative backgate voltage the electrostatic gating gives rise to a $V_{bg}$ independent $\eta$, which depends on the geometry. Similarly, at large positive $V_{bg}$ the same $\eta$ value is obtained but with bulk and edges that are electron doped (dashed lines). Near CNP a pronounced asymmetry in $\eta$ occurs due to the hole edge doping by $N_{imp}$. At small positive $V_{bg}$ the electrostatic electron doping of the edges compensates the hole edge doping due to $N_{imp}$, giving rise to a rapid drop in $\eta$ and in the corresponding $\mathcal{R}_{NL}$, consistent with the experimental observations in Fig. 1e. In this range of positive $V_{bg}$ the edges are hole doped while the bulk is electron doped, thus forming a $pn$ junction along the edges. The edge accumulation vanishes at a specific value of positive $V_{bg}$, which is $N_{imp}$ dependent, beyond which both the edges and the bulk become $n$-doped. Note that in this calculation we have ignored bulk disorder, $p_{dis} = 0$, which results in a diverging $\eta$ at CNP in Extended Data Fig. 3e because $p_b$ vanishes. Note also that the experimentally derived $N_{imp} \cong 0.18$ nm$^{-1}$ reflects the lower bound for edge accumulation since $\mathcal{R}_{NL}$ is governed by points of weakest charge accumulation along the edges. For comparison, Extended Data Fig. 3e also shows the case of pure electrostatic gating $N_{imp} = 0$ (green) with $p_{dis} = 2.9 \times 10^{10}$ cm$^{-2}$. In this case at CNP the transport is ohmic ($\eta = 0$) since $P_e$ vanishes at $V_{bg} = 0$ while $p_b = p_{dis}$ remains finite due to bulk disorder, and $\eta(V_{bg})$ is symmetric with respect to $V_{bg}$.

**Measurements and simulations parameters**

All the measurements were carried out at $T = 4.2$ K in an out-of-plane applied magnetic field $B$ up to 5 T.

**Fig. 1 and Extended Data Fig. 1.** Transport measurements acquired by conventional lock-in methods with *ac* voltage $V_0 = 9.0$ mV rms at $f = 7.566$ Hz applied to contact 1 using a low impedance voltage divider. The current was drained to ground at contact 2 and measured using a current preamplifier with input impedance of ~2 Ω. The nonlocal voltage was measured between contacts 3 and 4.

**Fig. 2. a-c**, Thermal imaging of the central part of the sample acquired using Pb SOT with effective diameter of 48 nm at a height of $h = 110$ nm above the top hBN surface at $B = 0$ T. An *ac* voltage $V_0$ was applied to contact 1 as in Fig. 1 but at $f = 66.66$ Hz. The temperature signal $T_{2f}$ was measured at frequency $2f$ as described in [32]. For each image $V_0$ was adjusted to keep the applied power $P = V_0 I_0 = 15$ nW constant; $V_0 = 6.1$ mV (**a**), 15.7 mV (**b**), and 4.7 mV (**c**). Image parameters: 290×100 pixels, pixel size 35.5 nm, 40 ms/pixel. **d-g**, Thermal imaging of the entire Hall bar using 100 nm diameter MoRe SOT at $h = 130$ nm at $B = 1$ T (**d-e**) and 5 T (**f-g**) and $P = 15$ nW. $V_0 = 28$ mV in **d,e** and 39.5 mV in **f,g**. Image parameters: 376×85 pixels, pixel size 46 nm, 40 ms/pixel.

**Fig. 3. a-f**, Scanning gate imaging using Pb SOT of Fig. 2 at $h = 110$ nm and $B = 5$ T. A *dc* voltage $V_0 = 5.5$ mV chopped by a square wave at a frequency of 25 Hz was applied to contact 1 while the current and the different contact potentials were measured simultaneously by several lock-in amplifiers. Image parameters: 163×110 pixels, pixel size 32 nm, 40 ms/pixel. **g-i**, Similar to **a-f** but scanning along the yellow line in **a** at $h = 30$ nm using $V_0 = 3.8$ mV. Image parameters: 217×125 pixels, pixel size 24 nm in $x$ and 80 mV in $V_{tg}$, 40 ms/pixel. **j-k**, COMSOL scanning gate simulation parameters: $B = 5$ T, $v_{tg} = 4$, and $\lambda = 30W$ ($V_{bg} = -0.2$ V, $\eta = 0.58$). The simulation was solved for each tip position within the 2D frame, with a step size of 50 nm in the $x$ direction and 25 nm in the $y$ direction. The edge disorder along the top edge in the frame was simulated by a spatially dependent edge accumulation $f(x)\frac{2P_e}{w}$, where $0 \leq f(x) \leq 1$, and $f(x) = 1$ along the rest of the edges. **l**, Line-scan simulation along the top edge with the same edge accumulation disorder $f(x)$ upon varying $v_{tg}$ from 0 to 3 in steps of 0.1 and tip position step size of 50 nm in $x$.

**Fig. 4. a-b,** COMSOL simulations with $B = 0$ T, $\eta = 0$. **c-d**, $B = 4$ T, $\mu = 2.6 \times 10^5$ cm$^2$V$^{-1}$s$^{-1}$, and $\eta = 0$. **e-f,** $B = 4$ T, $V_{bg} = -4$ V ($\eta = 0.05$, $\lambda = 2.54W$). **g-j**, $B = 4$ T, $V_{bg} = -4$ V, $v_{tg} = 2$.

**Supplementary Video 1**. Same parameters as in Figs. 2a-c at $B = 0$ T. For each frame $V_0$ was adjusted to keep the applied power $V_0 I_0 = 15$ nW with $V_0$ varying from 3.2 mV at $V_{bg} = -6$ V to 16.6 mV at $V_{bg} = 0$ V. The color scale is fixed. Near CNP the color is slightly saturated at a few points. Figures 2a-c show three frames from the movie with autoscaled color bar.

**Supplementary Video 2**. Same parameters as in Fig. 2f at $B = 5$ T. For each frame $V_0$ was adjusted to keep the applied power $V_0 I_0 = 15$ nW with $V_0$ varying from 8.8 mV at $V_{bg} = -10$ V to 39.5 mV at $V_{bg} = 0$ V. The color scale is fixed. Near CNP the color is slightly saturated at a few points. Figure 2f shows one frame from the movie.

**Supplementary Video 3**. Same parameters as in Supplementary Video 2 but with $V_{tg} = 8$ V. The color scale is fixed. Near CNP the color is slightly saturated at a few points. Figure 2g shows one frame from the movie.

**Supplementary Video 4**. Same parameters and setup as described above for Figs. 3a-f but with $V_0$ applied to contact 2 (instead of contact 1) and current drained at contact 1.

**Supplementary Video 5**. Same parameters and setup as described above for Figs. 3g-i.

**Supplementary Video 6**. COMSOL simulations vs. $B$ for $\eta = 0$ and 0.1 with $\mu = 2.6 \times 10^5$ cm$^2$V$^{-1}$s$^{-1}$. In regions of $\mathcal{J} > 1.2$ the red color is saturated.

**Supplementary Video 7**. COMSOL simulations for $B = 2$ T, $\mu = 2.6 \times 10^5$ cm$^2$V$^{-1}$s$^{-1}$, and backgate voltage ranging from $V_{bg} = -10$ V to $V_{bg} = 0$ V. In regions of $\mathcal{J} > 1.2$ the red color is saturated.

**Supplementary Video 8**. COMSOL scanning gate simulation with $B = 5$ T, $v_{tg} = 2$, $\lambda = 11.9W$ ($V_{bg} = -1$ V, $\eta = 0.2$), and uniform edge accumulation $f(x) = 1$. In regions of $\mathcal{J} > 1.2$ the red color is saturated.

**Supplementary Video 9**. COMSOL scanning gate simulation with nonuniform edge accumulation used for deriving Figs. 3j-k with $B = 5$ T, $v_{tg} = 4$, and $\lambda = 30W$ ($V_{bg} = -0.2$ V, $\eta = 0.58$). In regions of $\mathcal{J} > 1.2$ the red color is saturated.

**Captions of Supplementary Videos**

**Supplementary Video 1 | Thermal imaging of dissipation vs. $V_{bg}$ at $B = 0$ T.** Temperature maps $T_{2f}$ of the central part of the sample acquired with the scanning Pb SOT at $B = 0$ T and $V_{tg} = 0$ V with $V_{bg}$ varying from -6 V to 5 V with applied constant power $V_0 I_0 = 15$ nW. At large $|V_{bg}|$ most of the dissipation occurs at the top and bottom contacts outside the imaging frame, while dissipation within the sample is limited mainly to the central region where the current is expected to flow. On approaching CNP the dissipation extends from the central region into the right and left arms with enhanced thermal signal along the edges.

**Supplementary Video 2 | Thermal imaging of dissipation vs. $V_{bg}$ at $B = 5$ T and $V_{tg} = 0$ V.** Temperature maps $T_{2f}$ of the full Hall bar structure acquired with the scanning MoRe SOT at $B = 5$ T and $V_{tg} = 0$ V with $V_{bg}$ varying from -10 V to 10 V with applied constant power $V_0 I_0 = 15$ nW. At large $|V_{bg}|$ the dissipation is observed in the central region while near CNP the thermal signal extends over the entire sample. At $V_{bg}$ values corresponding to QH plateaus (e.g. $-8, -2, 3,$ and $8$ V) the dissipation in the sample is reduced with most of the dissipation occurring at the top and bottom contacts outside the imaging frame.

**Supplementary Video 3 | Thermal imaging of dissipation vs. $V_{bg}$ at $B = 5$ T and $V_{tg} = 8$ V.** Temperature maps $T_{2f}$ at $B = 5$ T and $V_{tg} = 8$ V with $V_{bg}$ varying from -10 V to 10 V with applied constant power $V_0 I_0 = 15$ nW. At large $|V_{bg}|$ the dissipation is observed in the central region while near CNP the thermal signal extends over the entire sample similar to Supplementary Video 2, but with enhanced signal along the edges. The tip potential $V_{tg}$ causes local depletion of the hole edge accumulation leading to diversion of the edge current into the bulk and to corresponding enhancement in dissipation as demonstrated by numerical simulations in Supplementary Video 9. The irregular pattern with enhanced thermal signal reveals the locations of suppressed edge charge accumulation along the edges.

**Supplementary Video 4 | Scanning gate microscopy vs. $V_{bg}$ at $B = 5$ T.** A constant $V_0 = 5.5$ mV is applied to contact 2 and $\mathcal{V}_{NL}$, $\mathcal{V}_3$ and $R_{2p}$ are measured as a function of the tip position in the central part of the sample (dotted area in Fig. 1a) with $V_{tg} = 8$ V at $B = 5$ T upon varying $V_{bg}$ from $-10$ V to 4 V. When the SOT is located above the sample edges the local depletion of the hole edge accumulation by the tip potential $V_{tg}$ causes suppression in $\mathcal{V}_{NL}$ and enhancement of $R_{2p}$ at $V_{bg}$ values corresponding to large $\mathcal{R}_{NL}$ in Fig. 1e. $\mathcal{V}_3$ is enhanced along the top edge and suppressed along the bottom edge in contrast to Fig. 3b due to $V_0$ applied to contact 2 instead of contact 1. The signals fade away at $V_{bg}$ values corresponding to QH plateaus.

**Supplementary Video 5 | Scanning gate microscopy along the top edge of the sample vs. $V_{tg}$ and $V_{bg}$ at $B = 5$ T.** A constant $V_0 = 3.8$ mV is applied to contact 1 and $\mathcal{V}_{NL}$ is measured as a function of the tip position along the top-right edge of the sample marked by the yellow line in Fig. 3a vs. $V_{tg}$ from $-2$ V to 8 V. Each frame corresponds to a different $V_{bg}$ that varies from $-10$ V to 4 V. A positive $V_{tg}$ depletes the hole edge accumulation causing suppression of $\mathcal{V}_{NL}$. This suppression is strongly position dependent due to edge accumulation disorder. Locations with weaker edge accumulation show a lower $V_{tg}$ threshold for $\mathcal{V}_{NL}$ suppression. Negative $V_{tg}$ causes additional hole accumulation along the edges which has no significant effect except at the weakest point of edge accumulation, which acts as a bottleneck for the edge current flow. At this point, a negative $V_{tg}$ can "repair" the suppressed edge accumulation, thus enhancing $\mathcal{V}_{NL}$ as observed at few values of $V_{bg}$. The overall $\mathcal{V}_{NL}$ signal fades away at $V_{bg}$ values corresponding to QH plateaus.

**Supplementary Video 6 | Simulations of nonlocal potential and current distributions vs. $B$ with and without edge accumulation.** Numerical simulations of the normalized potential $\mathcal{V}$ (top) and of the magnitude of the normalized current density $\mathcal{J}$ (bottom) upon increasing $B$ for $\eta = 0$ (left column) and $\eta = 0.1$ (right column). At low fields up to $B \cong 0.2$ T the normalized potential and the current density distributions for $\eta = 0$ and $\eta = 0.1$ evolve very similarly. At higher fields in absence of edge accumulation (left) the transport remains ohmic and essentially field independent. A small edge accumulation ($\eta = 0.1$), which has little effect at low fields, causes to a dramatic change in the potential and current distributions with increasing field (right). The potential becomes highly nonlocal and the spatial extent of the edge currents grows rapidly with field. At $B = 5$ T the nontopological edge currents extend all the way from the source (top contact) to the drain (bottom contact). In regions of $\mathcal{J} > 1.2$ the red color is saturated for clarity.

**Supplementary Video 7 | Simulations of nonlocal potential and current distributions vs. $V_{bg}$ at $B = 2$ T.** Numerical simulations of the normalized potential $\mathcal{V}$ (top) and of the magnitude of the normalized current density $\mathcal{J}$ (bottom) at $B = 2$ T upon varying $V_{bg}$ from $-10$ V ($\lambda = 0.57$W) to 0 V ($\lambda = 13.72$W). At small $\lambda$ the transport is similar to the ohmic regime. Upon increasing $\lambda$ the current along the edges expands, and the bulk current rotates it direction from vertical to horizontal, while the electric field rotates from horizontal to vertical. When the enhanced edge current extends all the way from source (top) to drain (bottom) the magnitude of the bulk current drops sharply. In regions of $\mathcal{J} > 1.2$ the red color is saturated for clarity.

**Supplementary Video 8 | Numerical simulation of scanning gate microscopy at $B = 5$ T.** Numerical simulations of the normalized potential $\mathcal{V}$ (top) and the normalized current density $\mathcal{J}$ (bottom) at $B = 5$ T and $V_{bg} = -1$ V ($\lambda = 11.9$W) upon scanning a positively biased tip (marked by a black circle) that causes local depletion of the hole carriers. When the tip resides in the bulk, the potential and the current are deformed only locally. When the tip is above the edge in the right-hand-side of the sample, the enhanced current flows along the edge from the source up to the tip position only, where it is diverted into the bulk with almost no current flowing along the rest of the edge. In the left-hand-side of the sample, the situation is opposite where the tip blocks the current from the source up to the tip position, whereas along the rest of the edge the current is gathered from the bulk into the edge and channeled towards the drain.

**Supplementary Video 9 | Simulations of scanning gate microscopy in presence of edge disorder.** Numerical simulations of the normalized potential $\mathcal{V}$ (top) and the normalized current density $\mathcal{J}$ (bottom) where the edge accumulation is position dependent along the top-right edge giving rise to a non-uniform potential. When the tip resides in the bulk, the potential and the current are deformed only locally. When the tip scans above the disordered top edge the suppression of the edge current is strongly position dependent with largest suppression occurring at the points of weakest edge charge accumulation. These types of simulations were used to produce Figs. 3j-l.